\documentclass[amsmath,amssymb,reprint,aip,floatfix]{revtex4-1}
\usepackage{siunitx}
\usepackage{graphicx}
\usepackage{dcolumn}
\usepackage{bm}
\usepackage{amsmath,amsfonts}
\usepackage{latexsym}
\usepackage{array}
\usepackage{adjustbox}
\usepackage[colorlinks=true,urlcolor=blue,citecolor=blue,linkcolor=blue]{hyperref}

\setcitestyle{round,authoryear,semicolon}

\begin{document}

\title[Accuracy of acoustic gunshot location]{Precision and accuracy of acoustic gunshot location in an urban environment}

\author{Robert B. Calhoun}
\email{rcalhoun@shotspotter.com}
\author{Clark Dunson}
\author{Murphey L. Johnson}
\author{Scott R. Lamkin}
\author{William R. Lewis}
\author{Robert L. Showen}
\author{Mark A. Sompel}
\author{Lester P. Wollman}
\affiliation{ShotSpotter, Inc., Newark, CA 94560, USA}

\begin{abstract}
The muzzle blast caused by the discharge of a firearm generates a loud, impulsive sound that propagates away from the shooter in all directions. The location of the source can be computed from time-of-arrival measurements of the muzzle blast on multiple acoustic sensors at known locations, a technique known as multilateration. The multilateration problem is considerably simplified by assuming straight-line propagation in a homogeneous medium, a model for which there are multiple published solutions. Live-fire tests of the ShotSpotter gunshot location system in Pittsburgh, PA were analyzed off-line under several algorithms and geometric constraints to evaluate the accuracy of acoustic multilateration in a forensic context. Best results were obtained using the algorithm due to Mathias, Leonari and Galati under a two-dimensional geometric constraint. Multilateration on random subsets of the participating sensor array show that 96\% of shots can be located to an accuracy of 15 m or better when six or more sensors participate in the solution.
\end{abstract}

\maketitle

\section{\label{Intro} Introduction}
The muzzle blast resulting from discharge of a firearm generates a loud, impulsive noise that propagates away from the shooter in all directions. The location of the shooter can be estimated by minimizing the difference between the measured and predicted arrival time of the muzzle blast on multiple time-synchronized sensors, a technique known as acoustic multilateration. Public safety agencies use networks of acoustic sensors to locate illegal gunfire in near real-time, 10--15 seconds after discharge \citep{aguilar2015gunshot, lawrence2018lessons, watkins2002technological, nlectc2007shotspotter}.

Acoustic data are also used in forensic context \citep{brustad2005a, maher2015advancing, maher2018audio, beck2019who, begault2019overview}. Acoustic multilateration cannot identify a shooter, but it can establish the discharge location and discharge time for each individual shot fired in a repeatable and objective manner. The efficacy and accuracy of an acoustic multilateration system depends on the number and position of reporting sensors, the accuracy to which the initial arrival time can be measured, and the validity of the propagation model, with the simplest acoustic propagation model assuming straight-line propagation. The present work outlines a procedure for computing shot locations from acoustic measurements of the muzzle blast using acoustic multilateration, and quantifies the precision and accuracy of the results obtained using data collected under controlled but realistic conditions.

The data derive from a series of live fire tests of the ShotSpotter gunshot location system conducted in Pittsburgh, PA. These tests were analyzed offline to identify the combination of algorithm and geometric constraint that offers the greatest precision when locating gunshots in a forensic context. Pittsburgh was selected as a study area for a number of reasons: first, a large number of tests were conducted (three firearms discharged from nine firing positions) covering neighborhoods with a density of structures similar to that of many ShotSpotter coverage areas; second, the hilly terrain offered the opportunity to investigate the importance of locating in three dimensions; and finally because the Pittsburgh array features an unusually high sensor density, allowing for simulation of the effects of deployed sensor density on detection rates and location accuracy. 

Source data from the Pittsburgh tests comparable to what is submitted as evidence in a criminal trial are provided as supplemental materials. These materials comprise time-stamped acoustic recordings, time- and location-stamped output of the pulse processing algorithms, and tagged pulse sets suitable for acoustic multilateration. These data permit other investigators to review or extend the current work.

\section{Theory of Gunshot Location}
\subsection{Source signal}
The muzzle blast of a gunshot is an impulsive acoustic source that emits power over a broad range of frequencies. \citep{weber} developed an expression for the pressure of spark explosions that has applicability to muzzle blasts; the resulting expression is similar to that introduced by \citep{friedlander1946diffraction}. The single parameter of the Weber model is the Weber radius $r_w$, the radius at which the supersonic combustion products decelerate to sonic velocity. Unlike explosions, muzzle blasts are highly directional \citep{maher2010directional,beck2011variations} but they can be modeled as a Weber source in which the size of $r_w$ depends on the angle between the gun barrel and the receiver \citep{hirschestimation}.

\begin{figure}[htpb]
\includegraphics[width={.45\textwidth}]{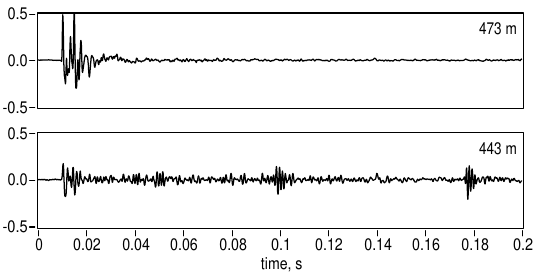}
\caption{Firearms discharged at similar ranges in an open environment (top: Kruger National Park, South Africa) and in an urban environment (bottom: city in the midwestern United States); the vertical scale is with respect to ADC full scale of 93 dB SPL = 1.0.}
\label{fig:signed}
\end{figure}

Muzzle blasts are simple impulses \citep{maher2009audio} and recordings made in open environments (such as Kruger National Park in South Africa, Figure \ref{fig:signed}, top) show reasonable agreement with simple models \citep{don1987impulse, embleton1996tutorial}. In built-up areas, propagation distances are shorter and the received waveforms are more complicated (Figure \ref{fig:signed}, bottom.) \citep{hornikx2016ten} notes that ``sound pressure levels in urban areas depend at one hand on the actual sound sources, and at the other hand on the propagation of sound from these sources in the environment'', a statement that is particularly applicable to urban gunshots, where a simple source signal becomes a complicated received signal as a result of the environment.  In urban environments the acoustic energy of the muzzle blast is transferred from source to a receiver via many different paths, including specular and diffuse reflection (echoes and reverberation), diffraction (transmission around structures), refraction (bending due to speed-of-sound gradient), as attenuated by absorption in the atmosphere and at surfaces \citep{kapralos2008sonel}. It is possible to model acoustic propagation in an urban environment using ray-tracing and other computational modeling techniques \citep{le2000comparison, parker2007acoustic, albert2010effect, mehra2014acoustic, pasareanu2018numerical, remillieux2012experimental, stevens2014spatial} but these models require a structural model of the area and are in some cases computationally demanding.

\begin{figure}[htpb]
\includegraphics[width={.45\textwidth}]{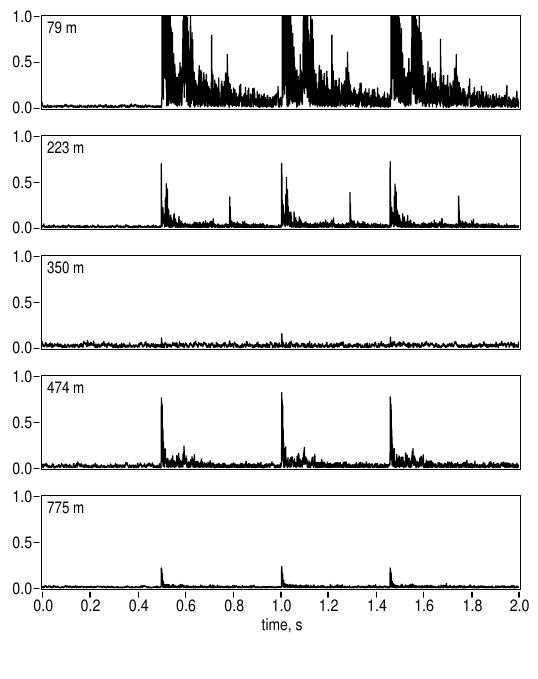}
\caption{Waveform envelopes for a sequence of 3 rounds from a 9mm pistol as recorded by sensors at \SI{79}{\meter}, \SI{223}{\meter}, \SI{350}{\meter}, \SI{474}{\meter} and \SI{775}{\meter} showing that range alone is not a good predictor of signal strength. Echo patterns vary from location to location, allowing the near line-of-sight paths to be identified through their greater self-consistency. The signals have been aligned in the time domain; the vertical scale is the same for all plots, with amplitude of 1.0 corresponding to ADC full scale of 93 dB SPL.}
\label{fig:hilbert}
\end{figure}

A simpler approach \citep{showen1997operational, showen2010complex} is to deploy more than the minimum number of sensors required for multilateration and assume a sufficient number of sensors will receive detectable levels of muzzle blast energy via paths well-approximated by a straight-line propagation model. These ``near line-of-sight'' pulses can be identified by constructing sets of pulses from appropriate sensors, computing a prospective location from the pulse set, and comparing arrival time predictions of the model with the measured arrival time on each sensor. The goal of this combinatorial optimization problem is to find the largest set of pulses that meet a mutual consistency test. Specular reflections (echoes) are rejected because the path between source and each sensor is distinct, resulting in different echo patterns in each received signal; only pulses from ``near line-of-sight'' paths pass the mutual consistency test. This principle is illustrated in Figure \ref{fig:hilbert}, which shows the envelopes of three successive shots detected on five different sensors. Strong echoes are detected on three of the sensors, but only the initial impulse from each shot is consistent with a straight-line propagation model on all the fives sensors. The mutual consistency method can only be used when the number of reporting sensors exceeds the minimum number of time differences of arrival (TDoAs) required for multilateration. The method works best when the sensor density is high enough to ensure gunshots are detected on at least two more sensors than required for multilateration, but not so high that the system generates false triggers on unwanted sources of impulsive noise such as construction noise.

\subsection{Pulse localization}
\label{pulselocalgo}

In built environments, the acoustic energy from a muzzle blast arrives via multiple paths. Energy transmitted via the shortest acoustic path arrives first; additional energy from the blast is delivered via reflections and other indirect paths over a period of time that may extend to several seconds. In the present work, \textit{signal} refers to the acoustic pressure change induced by the muzzle blast as measured by a remote microphone as a function of time; the goal of the pulse detection algorithm is to identify the arrival time of \textit{pulses}---fast rise time portions of the measured signal---that are plausibly due to muzzle blast energy arriving via the shortest acoustic path. Because shortest acoustic paths are often not line-of-sight paths, significant attenuation may occur and the acceptance criteria for muzzle blast impulses must necessarily be broad.

One effective strategy for detecting plausible muzzle blasts in built environments is to convolve the envelope of an acoustic signal with a step-detection kernel. This acknowledges that muzzle blast power will be broadly distributed in the time domain, while giving the most weight to the earliest-arriving signal.

Let $f(t)$ be the amplitude of an acoustic signal as a function of time $t$. The instantaneous magnitude (or envelope) of $f(t)$ is given by the magnitude of the analytic associate of $f(t)$
\[
\|f\| = \| f(t) + i \mathcal{H} ( f(t) )\|
\]
where $\mathcal{H} ( f(t) )$ is the Hilbert transform of $f(t)$ \citep{boashash2015time}. Define a weighted step-detection kernel $g(t)$ as
\begin{equation}
g(t) = 
     \begin{cases}
     t/\tau + 1 & \textrm{for\quad} -\tau < t \leq 0 \\\
        -1/2    & \textrm{for\quad}  0 <  t \leq \tau \\
        0       & \textrm{otherwise} \\
     \end{cases}
     \label{eqn:kernel}
\end{equation}
where $\tau$ is a length scale over which the signal should be averaged. 

Define $x(t)$ as the convolution the signal envelope $\|f(t)\|$ and kernel $g(t)$:
\begin{equation}
x(t) = (\|f\| * g)(t).
\label{eqn:finetime}
\end{equation}
The \textit{pulse arrival time} is defined as the time $t$ at which $x(t)$ is locally maximized. Pulses with values of $x(t)$ that fall below a threshold are discarded.

The kernel $g$ gives as much weight to the absence of signal prior to pulse arrival as it does to the amplitude of the signal that follows; this provides consistent estimates of the arrival time of the most direct (earliest-arriving) path, even when the energy received via that path is a small fraction of the total. Increasing the length scale $\tau$ reduces spurious triggering on echoes and reverberation, but an excessively long window reduces the ability to detect closely-spaced shots. Empirically, a value of \SI{50}{\milli\second} has been found to work well on both isolated shots and the range of semi-automatic (300 rpm) and fully automatic (450--950 rpm) firearms encountered in a public-safety context. Note this is 10--20 times the duration of a muzzle blast measured on an open range \citep{beck2011variations,maher2011acoustical,routh2016recording}, which is on the order of \SI{2}{\milli\second}.

\begin{figure}[htpb]
    \centering
\includegraphics[width={.45\textwidth}]{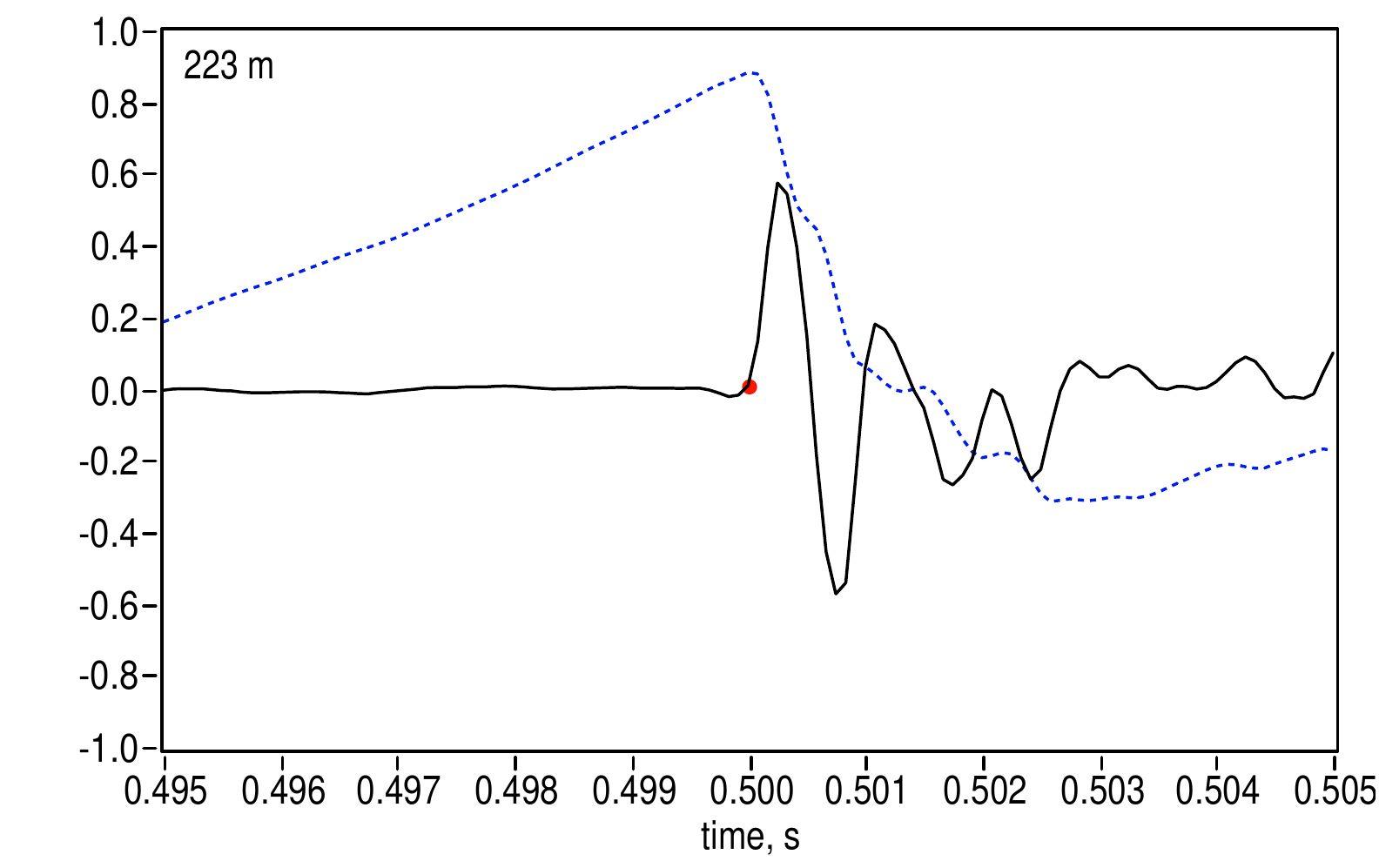}
    \caption{Pulse arrival time (red dot) plotted on the original waveform for sensor at \SI{223}{\meter} distance from Figure \ref{fig:hilbert}. Dotted line shows output of pulse localization function $x(t)$ from \eqref{eqn:finetime}. The full scale amplitude approximately 93 dB SPL.}
    \label{fig:finetime}
\end{figure}

\subsection{Pulse Selection}
In urban environments, true line-of-sight paths are rare. ``Near line-of-sight'' paths, where the received signal is line-of-sight after a single diffraction over a barrier, can be used for location because the typical height of buildings is small compared with the distances between acoustic sensors. Use of non-line of sight paths with significant time delays, such as those generated by specular reflections from buildings and other hard surfaces, will cause significant location error if used in a multilateration calculation. Associating pulses from one shot with pulses from a different shot will also result in large multilateration error. 

The construction of sets of pulses that are appropriate for multilateration becomes more difficult as the number of shots and number of shooters increases. ShotSpotter's pulse selection algorithms use a combination of combinatorial optimization, pattern-matching and numerical optimization methods to search for the largest possible set of mutually-consistent pulses for each shot. The selection of appropriate pulses for multilateration is an involved topic \citep{calhoun2011multishot, showen2010complex, beck2019who} and a full discussion of the methods used is outside the scope of the present work. To facilitate review of the present multilateration results, the output of the ShotSpotter pulse selection routines (comprising sets of acceptably ``near line of sight'' pulses for each shot) are provided as part of the supplemental materials, along with sufficient audio data to verify that the selected pulses are associated with the shortest acoustic path. (Supplementary materials  containing audio files, pulse properties, and tuples of near line-of-sight pulses for each shot that are suitable for multilateration are currently available at \url{https://github.com/ShotSpotter/research.accuracy-of-gunshot-location}.) Because the search for pulse sets is not exhaustive, the tuples provided are not guaranteed to be an optimal subset for each test. However, all pulses in a tuple will have an error of less than \SI{40}{\milli\second} between the measured arrival time and the predictions of a straight-line propagation model.

\subsection{Source Localization}
\label{multilateration}
Acoustic multilateration, the location of a source from multiple time difference of arrivals (TDoAs), has been in use since World War I when French \citep{aubin2014war}, Canadian \citep{finan1997mcnaughton} and British forces \citep{van2005lawrence} used sound ranging to locate the source of artillery fire. Many multilateration algorithms have been documented since that time. A multilateration solution in $d$ dimensions requires at least $n = d + 1$ reporting sensors. Some $n = d + 1$ solutions have two mathematically consistent solutions. Additional methods not discussed in the present work, such as measurements of angle-of-arrival (AoA) or received signal power, are required to resolve ambiguous solutions, so the present work focuses on solutions for which $n \geq d + 2$, for which the multilateration solution is always unique given non-degenerate sensor placement.

The TDoA source location problem reduces to finding the point of intersection of multiple hyperboloids, so the algorithms comprise variations on a common theme. ShotSpotter uses multiple algorithms because they exhibit different numerical stability and sensitivity to the choice of a reference sensor. Below we analyze the accuracy of four algorithms and three constraints under a variety of array densities to compare their relative performance under real-world test conditions. A goal of the present work is to identify which algorithm is best-suited for forensic work. All solutions discussed assume line-of-sight propagation in a homogeneous, stationary medium.

\subsubsection{Reddi}
This implementation is based on \citep{reddi}. With reference sensor at $(0,0,0)$ and other sensors at $(x_i,y_i,z_i)$, $i=1,2,3, \dots n$, source location $(x_*,y_*,z_*)$ can be found from relative arrival times $t_i$ by equating the squared distance from the reference to the source with the square distance of the reference to the other sensors:
\begin{align}
c t_i   = S_i - \sqrt{x^2 + y^2 + z^2} 
\end{align}
for $i = (1,2, \dots n)$ where
\[
S_i^2 = (x - x_i)^2 + (y - y_i)^2 + (z - z_i)^2.
\]
This reduces to a quadratic equation that can be solved for source location $(x_*,y_*,z_*)$. Discharge time $t_*$ is obtained from the mean value of $t_i - S_i / c$ at $(x_*,y_*,z_*)$. This implementation requires a minimum of $d+1$ TDoAs, where $d$ is the number of dimensions.

\begin{figure}
    \centering
\includegraphics[width={.45\textwidth}]{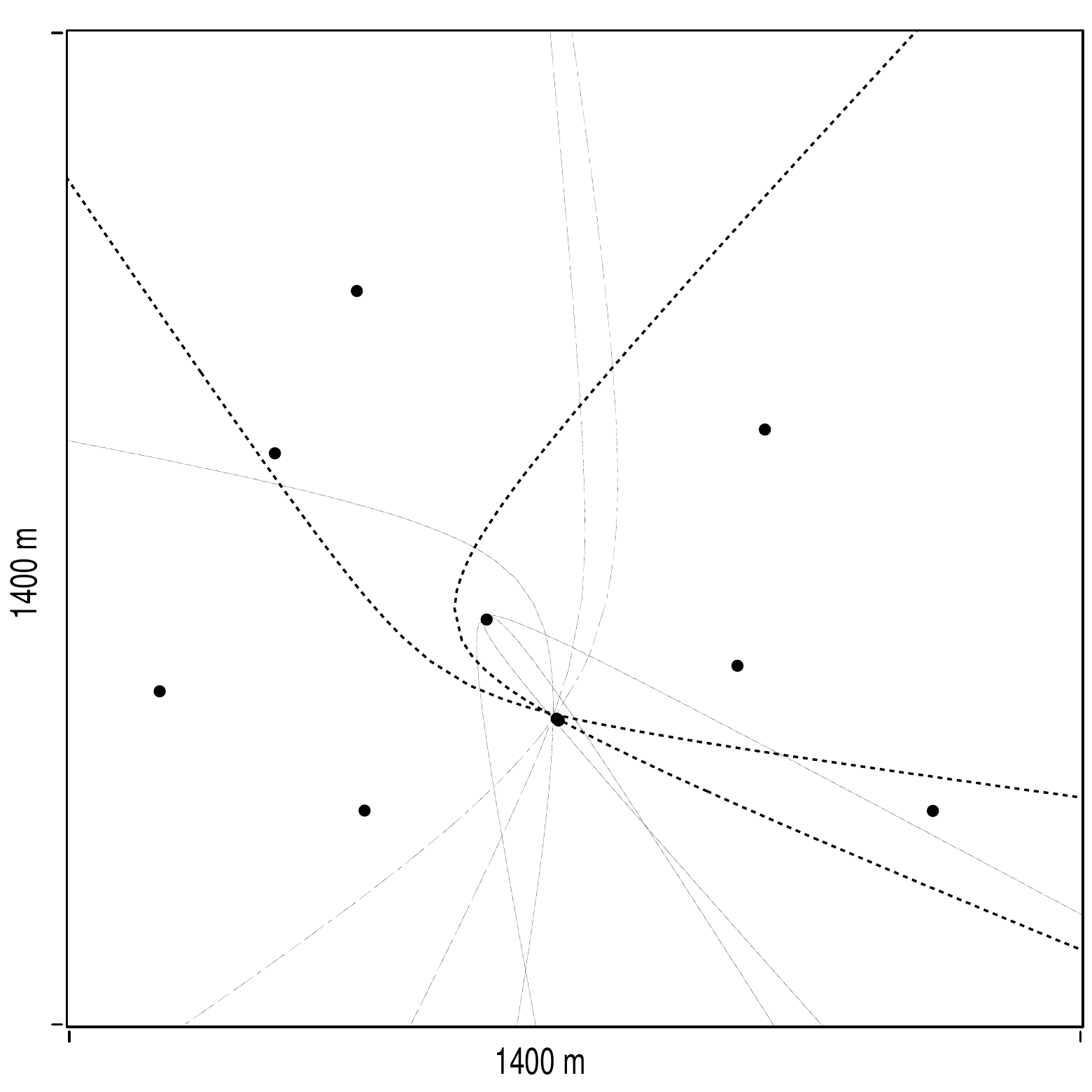}
    \caption{Graphical representation of the intersecting hyperbola approach characteristic of Reddi algorithm. The reference sensor that defines the time differences of arrival is the first reporting sensor. The sample data is from test 1 at Firing Position 4.}
    \label{fig:hyperbolas}
\end{figure}

\subsubsection{MLG}
\citep{mathias2008efficient} avoid use of a reference sensor by minimizing the squared error on each sensor. ShotSpotter's ``MLG'' implementation slightly modifies this algorithm by solving for time-distance $(c t)$ instead of $t$, which improves computational stability. For the simplified case of overdetermined systems ($n > d+1$) with receiver matrix
\begin{equation}
\label{eqn:receivermatrix}
\mathbf{R}=
\begin{bmatrix}
  x_0 & x_1 & \dots & x_i \\
  y_0 & y_1 & \dots & y_i \\
  z_0 & z_1 & \dots & z_i \\
  -c t_0 & -c t_1 & \dots & -c t_i \\
  -c/2 & -c/2 & \dots & -c/2 \\
\end{bmatrix},
\end{equation}
the solution vector is given by
\begin{equation}
\label{eqn:mlgsolution}
\begin{bmatrix}
x \\ 
y \\
z \\ 
c t \\
v/c \\
\end{bmatrix} 
= 
\frac{1}{2}
\mathbf{R}^\dagger 
\begin{bmatrix}
x_0^2 + y_0^2 + z_0^2 - (c t_0)^2\\
x_1^2 + y_1^2 + z_1^2 - (c t_1)^2\\
\dots  \\
x_i^2 + y_i^2 + z_i^2 - (c t_i)^2\\
\end{bmatrix}
\end{equation}
where $v$ is an error term and $\mathbf{R}^\dagger$ is the pseudoinverse of $\mathbf{R}$. See \citep{mathias2008efficient} for handling of the fully determined case ($n = d+1$) , and for a quantification of error propagation in multilateration.

\subsubsection{LeastSquares}
The ``LeastSquares'' implementation is another least-squares minimizer, based on time difference of arrivals between pairs of sensors. Equating the squares of the distance and the time-distance gives one equation for each sensor $i$:
\begin{equation}
(x_i - x)^2 + (y_i - y)^2 + (z_i - z)^2 = (c t_i - c t)^2.
\label{eqn:distancetimedistance}
\end{equation}
Subtracting the $i$th equation from the ${(i+1)}$-st eliminates quadratic terms. Define $X_i$ to be the difference between the LHS (distance) and RHS (time-distance) of the $i$th difference equation. When
\[
\chi^2 = \sum_{i=1}^{n-1} X_i^2,
\]
the minimum value of $\chi^2$ is the point at which
\[
\frac{\partial \chi^2}{\partial x}   =  
\frac{\partial \chi^2}{\partial y}   =
\frac{\partial \chi^2}{\partial z}   =
\frac{\partial \chi^2}{\partial c t} =  
0.
\]
This gives a linear system in $d+1$ variables:
\begin{equation}
\begin{bmatrix}
a_i a_i & a_i b_i & a_i c_i & a_i d_i\\
b_i a_i & b_i b_i & b_i c_i & b_i d_i\\
c_i a_i & c_i b_i & c_i c_i & c_i d_i\\
d_i a_i & d_i b_i & d_i c_i & d_i d_i\\
\end{bmatrix}
\begin{bmatrix}
x \\
y \\
z \\
c t \\
\end{bmatrix}
=
\begin{bmatrix}
a_i e_i \\
b_i e_i \\
c_i e_i \\
d_i e_i \\
\end{bmatrix}
\label{eqn:lsloc}
\end{equation}
where
\begin{align}
a_i & =  (x_{i+1} - x_i), \nonumber \\
b_i & =  (y_{i+1} - y_i), \nonumber  \\
c_i & =  (z_{i+1} - z_i), \nonumber  \\
d_i & =  -c (t_{i+1} - t_i), \: \textrm{and} \nonumber \\
e_i & =  \dfrac{1}{2} (
x_{i+1}^2 + y_{i+1}^2 + z_{i+1}^2 + c^2 t_i^2 - x_i^2 - y_i^2 - z_i^2 - c^2 t_{i+1}^2 ) \nonumber .
\end{align}

This requires an additional TDoA compared with the Reddi and MLG implementations.

\subsubsection{Iterative Discharge Time}
Combustion of the propellant used firearms generates an optical signal known as the \textit{muzzle flash} that is strongest in the medium-wave IR region \citep{pauli2004infrared}. If the discharge time $t_*$ is known from muzzle flash timing \eqref{eqn:distancetimedistance} reduces to finding the point of intersection of circles (or spheres) of known radius $c(t_i - t_*)$. This is a simplified version of \eqref{eqn:lsloc}.

\begin{figure}
    \centering
\includegraphics[width={.45\textwidth}]{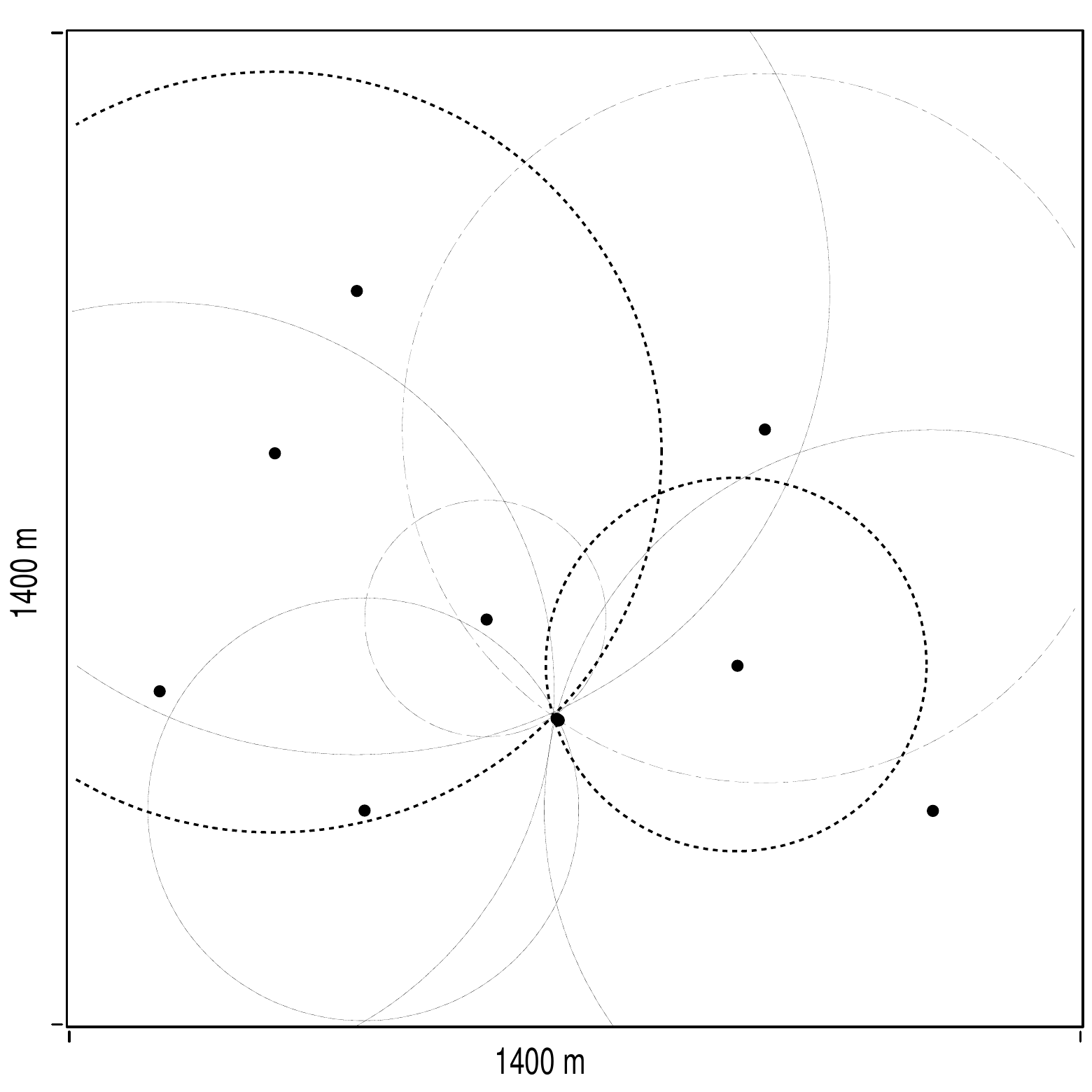}
    \caption{Graphical representation of the intersecting circle approach characteristic of Iterative Discharge Time algorithm. The sample data is from test 1 at Firing Position 4.}
    \label{fig:circles}
\end{figure}

When no muzzle flash time is available the discharge time can be determined iteratively. The ``Iterative Discharge Time'' (IDT) implementation uses simplex gradient descent to find $t_*$ based on starting value of $t_* = t_0 - 1.0$, where $t_0$ is the arrival time on the nearest (earliest) sensor and the constant roughly equal to the mean distance between a shooter and the first reporting sensor. The main advantage over pure TDoA algorithms is that the range can be bounded, which is helpful when locating sources outside the sensor array.  An additional benefit of all intersecting circle solutions in a forensic context is that the location tuple obtained $(x_*, y_*, t_*)$ can readily be checked using the standard tools of plane geometry, a ruler and a compass. A graphical representation of an intersecting circles solution is shown in Figure \ref{fig:circles}.

\subsection{Geometric Constraints}
Acoustic multilateration can be done in two or three dimensions. For two-dimensional multilateration, the sensor positions are projected against a plane and the multilateration solution returned is computed in this plane. At least three TDoAs are required. For three-dimensional multilateration, the sensor elevation data is used without modification and the multilateration solution computed in three-dimensional space. Four TDoAs are required for multilateration in three dimensions.

There are two potential advantages of three-dimensional multilateration. First, a 3D straight-line distance propagation model can be used. This is potentially more accurate. Second, the elevation of the shooter can be estimated. The downside of 3D multilateration is that the additional degree of freedom allowed by three-dimensional multilateration can lead to less accurate 2D results if the sensor position matrix is poorly conditioned, e.g. the range of vertical elevations of reporting sensors is small compared with the horizontal spacing between them. With poorly-conditioned arrays, small errors in the measurements of arrival time, sensor position, or speed of sound may result in large changes in the shooter's putative elevation.

There are plausible use cases for detecting gunshots above ground level, including locating shots fired from the upper floors of a structure and assisting in the classification of aerial fireworks. Nevertheless the vast majority of firearms are discharged at ground level, so a practical approach is to do the bulk of pulse selection in two dimensions, and then refine the solution using a 3D model when sufficient data are available.

\begin{figure}[htpb]
    \centering
\includegraphics[width={.45\textwidth}]{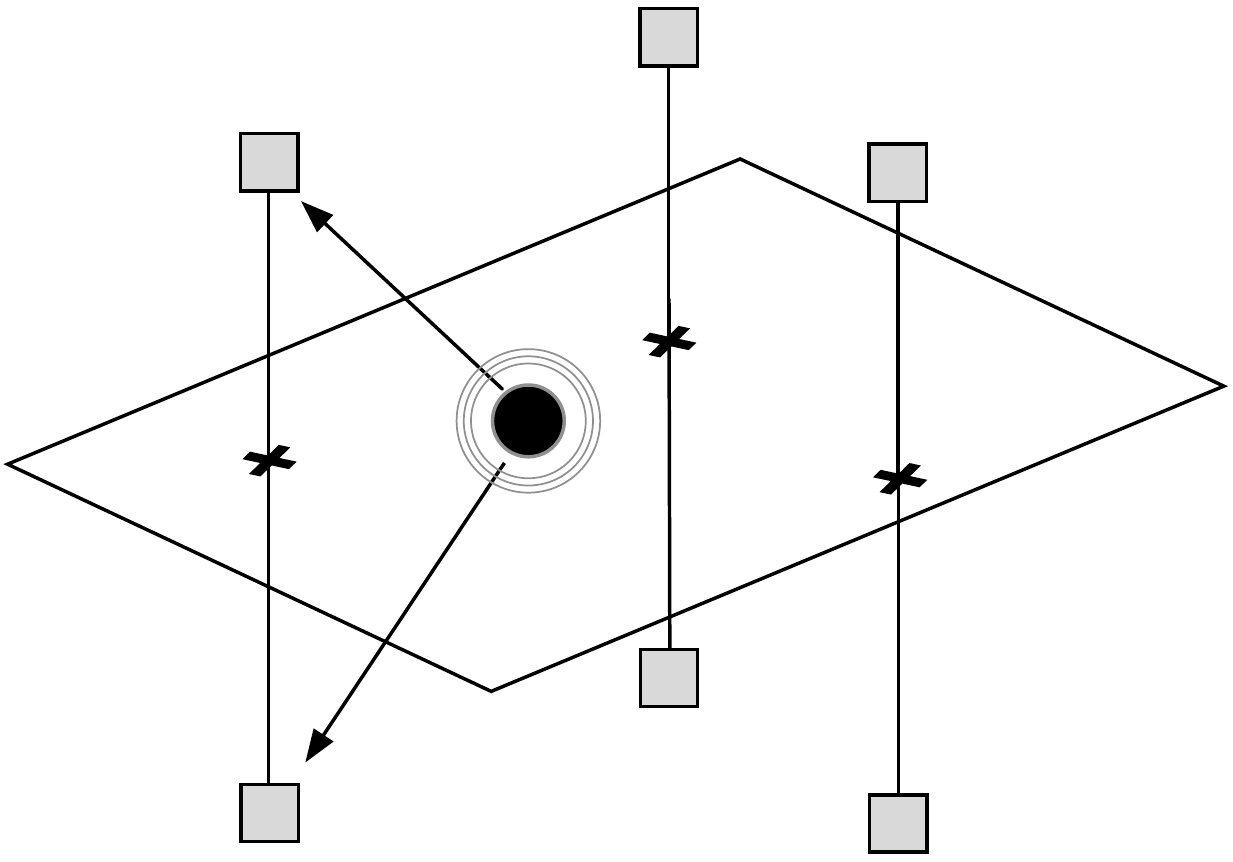}
    \caption{Method for constructing a sensor array through a ground plane by reflection. This permits use of elevation data from the remote sensors while constraining the output to the specified plane.}
    \label{fig:reflectedsensors}
\end{figure}

An approach that enables use of a 3D propagation model while avoiding the instability of a full 3D solution is to constrain the shooter location to a two-dimensional plane. This ``2.5D'' method \citep{calhoun2013} refines an initial 2D solution by assuming the shooter is standing on the ground (or on the roof of a building) and constructing $\mathbf{R}$ with a duplicate set of sensors with identical arrival time but position reflected through the horizontal plane containing the shooter, Figure \ref{fig:reflectedsensors}. The method requires use of a digital elevation and/or building model to provide an estimate for $z_*$, the $z\textrm{-elevation}$ of the shooter. With ground plane normal to $\hat{z}$, each TDoA tuple $(x_i, y_i, z_i, t_i)$ in \eqref{eqn:receivermatrix} is supplemented with a second TDoA tuple $(x_i, y_i, 2 z_*  - z_i, t_i)$. This doubles the number of columns in \eqref{eqn:receivermatrix} and the number of rows in \eqref{eqn:mlgsolution} but no changes in algorithm are required. The solution will be constrained to elevation $z_*$ by the symmetry of the input data.

\subsection{Weather Effects}
Atmospheric conditions affect sound propagation and therefore the accuracy of an acoustic gunshot location system. The speed of sound varies only slowly with temperature, and little with humidity or pressure, so the dry air relationship for the speed of sound $c$ may be used with acceptable accuracy. This is:
\begin{equation}
c = 20.03 \sqrt{T + 273.15} \quad \si{\meter\per\second}    
\end{equation}
where $T$ is the temperature in \si{\celsius}.

Sound energy is transmitted as a compression wave in a fluid; a moving fluid carries the sound energy with it. The multilateration routines of Section \ref{multilateration} assume a stationary medium, but the effects of a homogeneous moving fluid can be incorporated by shifting the apparent position of each sensor to account for air motion from when the shot is fired to when it is detected. With the $z$-axis component of wind vector $\vec{v}$ assumed to be zero, the wind-corrected position of the sensors is given by:
\begin{equation}
    \begin{bmatrix}
    x_i^{\prime} \\
    y_i^{\prime}
    \end{bmatrix}
    =
    \begin{bmatrix}
    x_i \\
    y_i
    \end{bmatrix}
    -
    (t_i - t_*)
    \begin{bmatrix}
    v_x \\
    v_y
    \end{bmatrix}
\end{equation}
where $t_i$ the arrival time at each sensor and $v_x$ and $v_y$ are the $x$- and $y-$components of the wind vector. The discharge time $t_*$ is not known \textit{a priori} but it can be estimated by subtracting an appropriate value from the arrival time at the first reporting sensor. A suitable value is the (historic) mean distance from the shooter to the first reporting sensor; a value of \SI{1.0}{\second} is appropriate for arrays with sensor densities in the range of \SIrange{2}{12}{{sensors}\per\kilo\meter\squared}. The wind-correction process can be repeated with a more accurate value after initial computation of discharge time $t_*$.

\begin{figure}
    \centering
\includegraphics[width={.45\textwidth}]{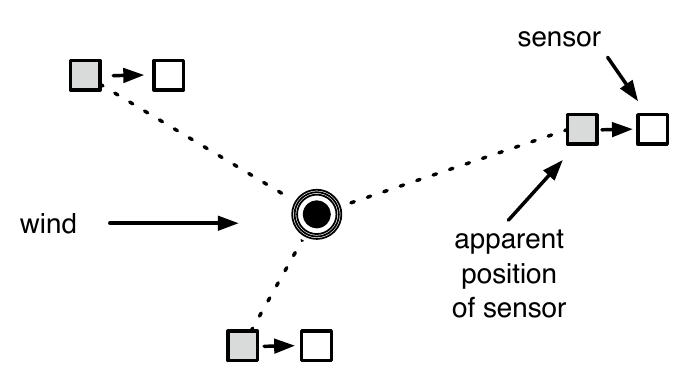}
    \caption{Adjustment of sensor positions to take homogeneous fluid movement into account. Energy is propagated from source to sensor via a combination of bulk fluid flow and acoustic propagation. The effects of bulk fluid flow can be incorporated by shifting the apparent position of the sensor to the position that would have received the signal in the absence of fluid flow.}
    \label{fig:wind}
\end{figure}

Other deleterious effects of wind are not so easily handled. Wind increases the background noise detected by outdoor microphones \citep{raspet2006framework}, an effect that can be ameliorated but not entirely eliminated with proper windscreens \citep{hosier1979microphone, lin2014effect, schomer1990methods, wuttke1992microphones}. Wind noise reduces the signal-to-noise ratio of gunshot impulses, making them harder to detect. Gradients in wind speed and temperature with elevation refract the sound, bending it upwards in the upwind direction and downwards in the downwind direction \citep{wiener1959experimental, wilson2002windref, franke1988numerical}. Upwind refraction creates a ``shadow zone'' where the gunshot signal is not detectable. Because of the potential lack of participation from upwind sensors during windy conditions, additional sensors must be to deployed to enable successful multilateration in all weather conditions. The ShotSpotter system does not otherwise compensate for the effects of wind or temperature gradients.

\section{System Design}
\label{sysdesign}
\subsection{Array design}
A typical deployment density is \SIrange{2} {12}{{sensors}\per\kilo\meter\squared}, with the density dependent on the density of structures and foliage in the covered area. The low end of this range is used in open areas where few structures impede acoustic propagation, such as anti-poaching deployments; higher sensor densities are used in areas with a high density of structures, high background noise (traffic, rapid transit systems), difficult wind conditions, or unfamiliar environments. Sensors are installed on elevated locations such as rooftops and utility poles to increase the number of near line-of-sight audio paths, and to minimize background noise. Wind noise is mitigated with a \SI{3.5}{\centi\meter} layer of open-cell foam around the microphones.

\subsection{Sensor processing}
\begin{figure}[htpb]
    \centering
\includegraphics[width={.45\textwidth}]{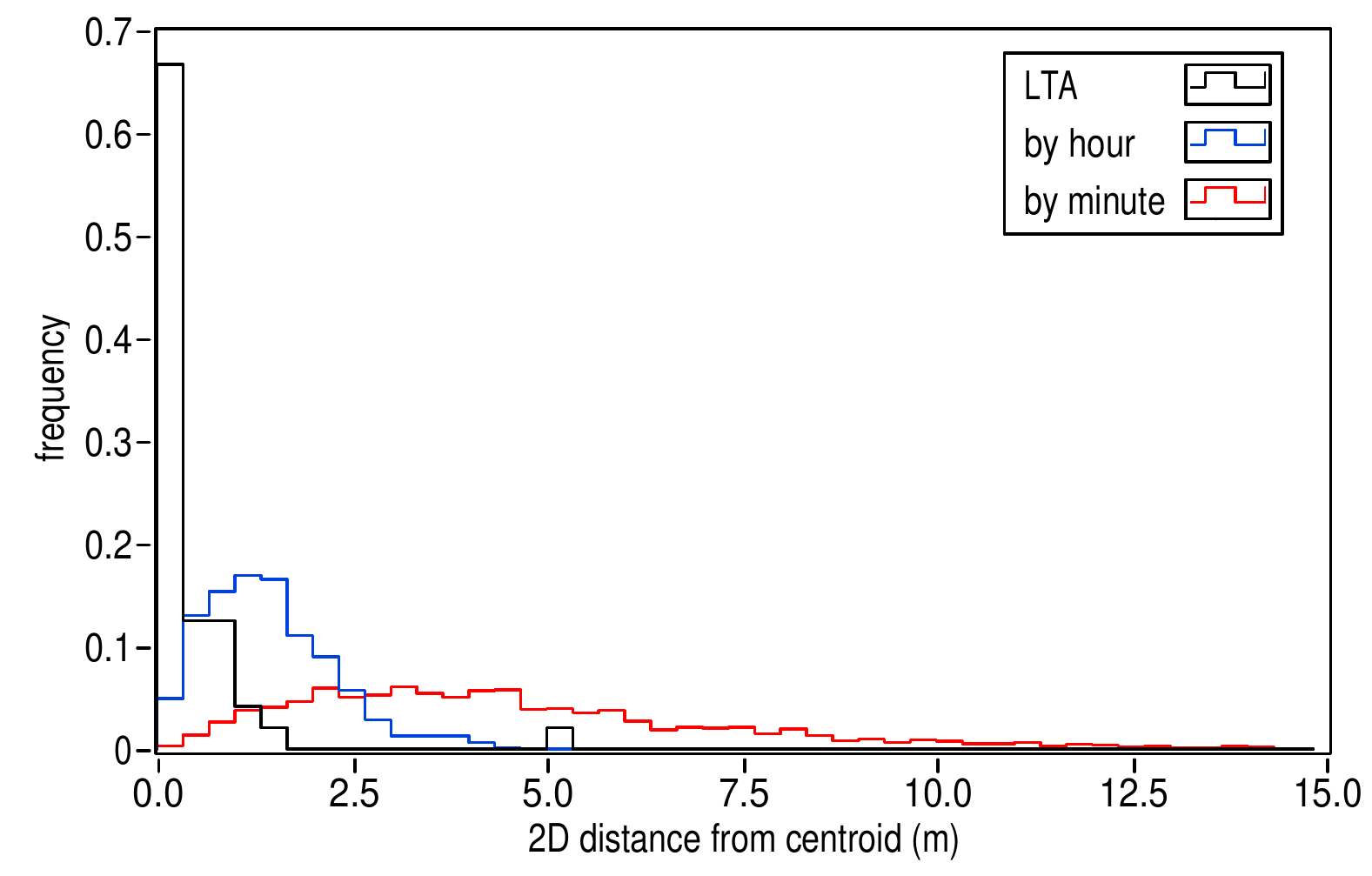}
    \caption{Horizontal GPS performance of a typical stationary sensor, showing the distance from the centroid of GPS measurements sampled once per minute, of the hourly averages, and of the long-term average of hourly values. 50\% CEP is slightly worse than predicted by the datasheet but averaging reduces the RMS distance from the centroid to less than \SI{0.02}{\meter}.}
    \label{fig:gpshisto}
\end{figure}

All signal processing and pulse detection operations are performed in the field on remotely-deployed single-board computers based on the Texas Instruments AM335x platform. Sensor position is provided by a u-blox MAX 7 GPS chipset, and the Linux system clock is disciplined using the PPS signal from the GPS. The u-blox MAX-7 series of GPS chipsets have a datasheet location accuracy of 2.5m (50\% circular error probability (CEP)) and timepulse accuracy \SI{30}{\nano\sec} (RMS). Actual performance depends on the satellite constellation visible from the installed sensor position. Because sensors are rarely moved, averaging over long periods of time is a feasible method for improving location precision. Each hour, the mean position over the previous hour is stored, and a long-term average position re-estimated based on values from up to 180 days of hourly records, (Figure \ref{fig:gpshisto}) with the multilateration calculations making use of the long-term averages. It is expected that the increasing availability of high-accuracy real-time kinematic (RTK) GPS chipsets will reduce or eliminate the need for position averaging in future multilateration systems.

Audio sound pressure level is measured using Kingstate KECG2742WBL waterproof electret microphones and digitized using a Cirrus Logic WM8737 analog-to-digital converter, which samples 24-bit audio data at 48 kHz. Microphone sensitivity is $-42 \pm 3$ dB SPL, giving a full-scale digital output of approximately 93 dB SPL. Network connectivity is provided by an on-board cellular data module.

For performance reasons, the pulse localization algorithm described in \ref{pulselocalgo} is implemented as a discrete-time algorithm. The digitized input stream is split up into 2048-length chunks with 50\% overlap so that a Hann window can be applied without data loss. After windowing, data is converted into the frequency domain for filtering and analysis. A typical pulse localization response is shown in Figure \ref{fig:finetime}.

\begin{figure}[htpb]
    \centering
\includegraphics[width={.45\textwidth}]{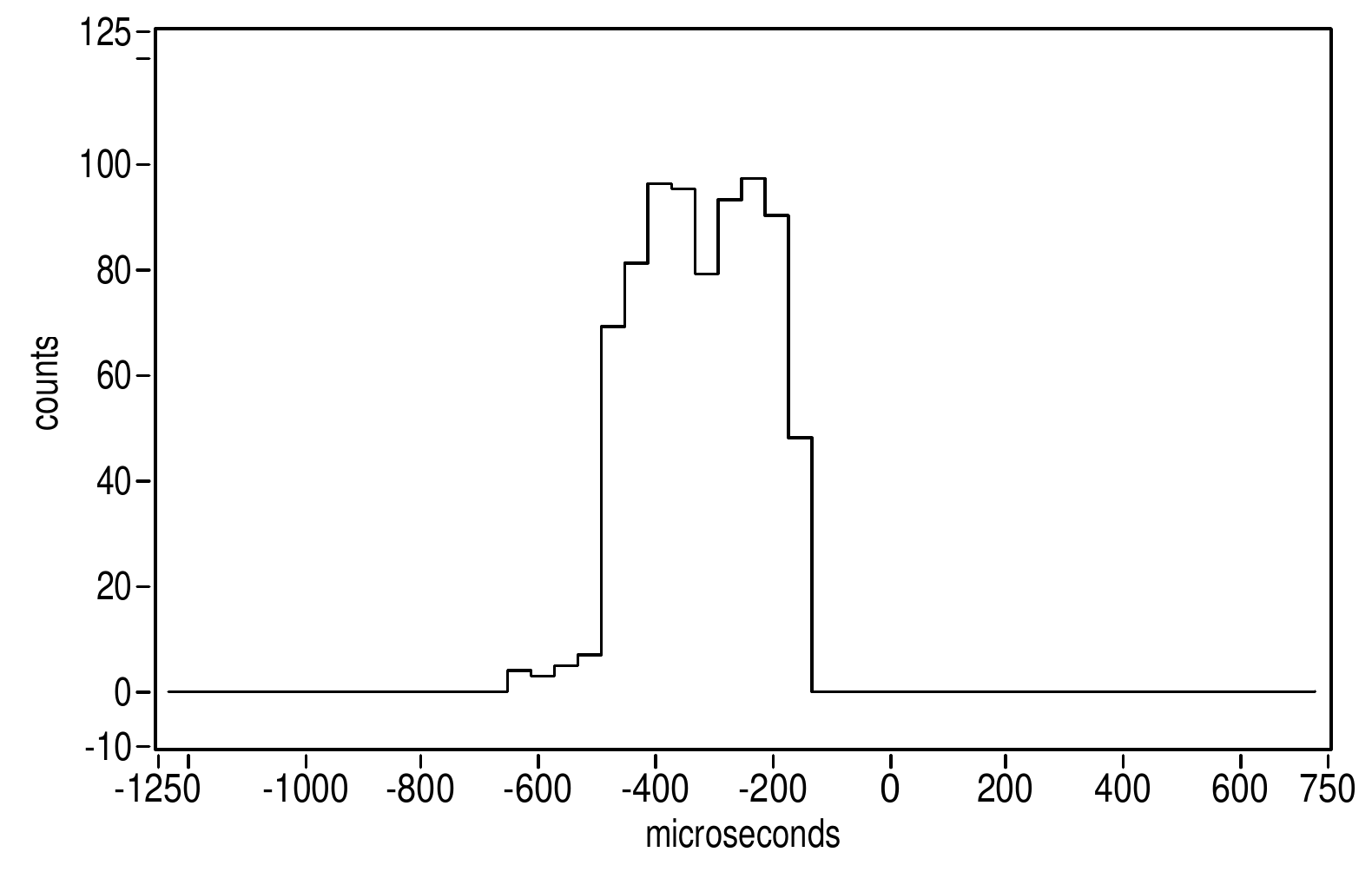}
    \caption{Pulse start timing measurements using simulated gunshot impulses triggered by GPS timepulses}
    \label{fig:timingdata}
\end{figure}

\begin{table}[htbp]
    \centering
\begin{ruledtabular}
\begin{tabular}{ccc} 
system & $\sigma$ (\si{\micro\second}) \\
\hline
GPS timepulse & \num{3 e-2} \\
Linux system clock & \num{1 e-3} \\
Pulse start measurement & \num{1 e2} \\
\end{tabular}
\end{ruledtabular}
\caption{Scale of contributions to pulse arrival time error}
\label{tab:timingerror}
\end{table}
Pulse arrival time measurements are derived from audio synchronized to the GPS PPS timepulse. As shown in Table \ref{tab:timingerror}, the timing errors associated with the GPS system are negligible compared with the uncertainty in pulse arrival time measurement. The net effect of all timing errors is estimated by running the pulse detection algorithm on gunshot-like impulsive signals synchronized with GPS PPS. This measurement takes into account multiple sources of error. The measured timing error through the ShotSpotter sensors has a mean of $\mu = \SI{-316}{\micro\second}$ with a standard deviation of $\sigma=\SI{99.6}{\micro\second}$, with a typical measurement shown in Figure \ref{fig:timingdata}. The pulse arrival time data provided and used for multilateration in the current work are to millisecond precision.

The audio processing code analyzes the first \SI{250}{\milli\second} following pulse arrival time to determine the likelihood that the impulse is due to a gunshot muzzle blast. Measured characteristics of the impulse include the integrated power of the impulse, the shape of its envelope, and a normalized power spectrum. These data are transmitted to a common endpoint for the multilateration computation.

\subsection{Classification}
\begin{figure}[htbp]
    \centering
\includegraphics[width={.45\textwidth}]{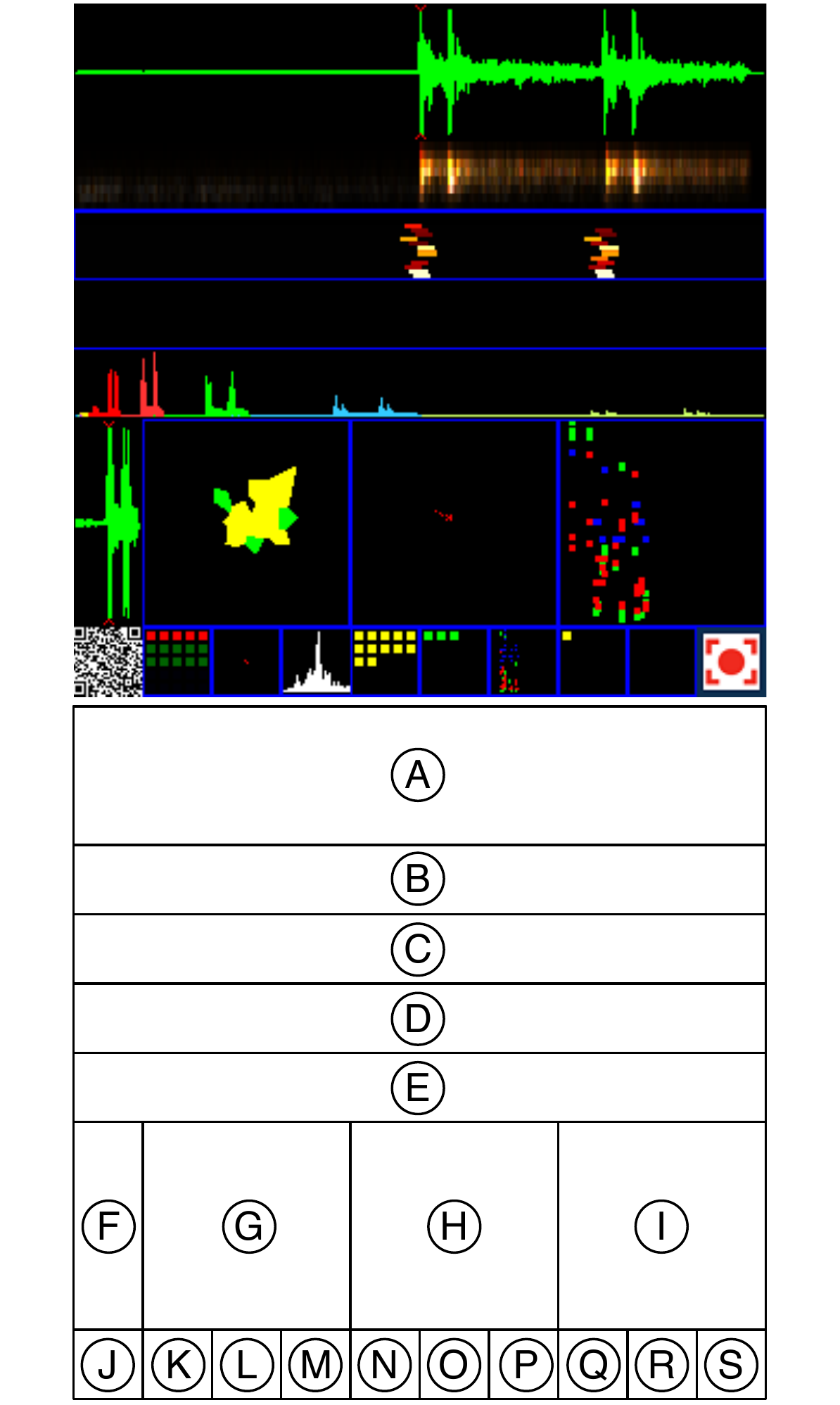}
    \caption{The top image shows a typical mosaic image used for machine classification with ResNet image classifier. This example is from a \SI{9}{\milli\meter} firearm discharged at Firing Position 8. Each image used for classification is comprised of 19 individual tiles; some tiles are derived from a 2-second subset of 12 kHz audio, while others are derived from the characteristics of the incident as a whole. See text for tile descriptions.}
    \label{fig:resnet}
\end{figure}

The original 48 kHz data is discarded after pulse processing; data downsampled to 12 kHz is cached on the sensor for 30 hours to support forensic analysis. Following detection of a potential gunshot, a 2-second subset of downsampled (12 kHz) FLAC-encoded audio is retrieved from the two closest sensors and used to build an \textit{image mosaic} (Figure \ref{fig:resnet}) comprising features characteristic of the gunshot audio on each sensor and features characteristic of the incident as a whole. A full description of image mosaic construction is outside of the scope of the present work, but the tiles comprising the image can be summarized as follows: A) waveform image; B) discrete wavelet spectrogram of the de-noised waveform; C) pulse time-of-arrival markers aligned using the shooter-sensor time delay computed via multilateration; D) discrete wavelet spectrogram of the noise (complement of tile B above); E) wavelet transform of the first impulse; F) waveform of the first impulse; G) participating area, as defined by the Voronoi diagram of participating sensors; H) location of recent nearby incidents I) plots of various per-sensor parameters as a function of distance; J) version-tracking code; K) time-of-date counters; L) individual shot locations; M) power spectrum; N, O, P) pulse feature counters; Q, R) recent incident counters; S) logo. Discrete wavelet transforms are performed using the biorthogonal 2.2 wavelet and scaling functions.

Classification of the mosaic images is performed using the ResNet \citep{he2016deep} image classifier trained using field-collected data labelled by human reviewers. Each mosaic image (one from the nearest sensor, and one from the second-nearest) is classified independently; the outputs of the two classifiers are combined to minimize the likelihood of false negatives. In the vast majority of cases, ``ground truth'' (e.g. shell casings, video recordings) is not available, and it is to be expected that some training data are misidentified.

Incidents classified as gunfire are pushed to the public safety agency's dispatch system and/or mobile devices. All of the pulse data associated with an incident is stored in a database so that the measurements made by the remote sensor are available for later analysis. If a forensic report is requested (typically in the course of a criminal investigation) within 24 hours, eight seconds of 12 kHz downsampled audio may be retrieved from sensors participating in the incident. The downsampled audio is invaluable for verifying that the pulse selection algorithms have grouped the input pulse data into appropriate subsets for multilateration.

\subsection{Weather Measurements}
Reliable temperature data with a precision of \SI{+-1}{\celsius} are available via NWS METAR readings from major airports; measurements from Pittsburgh International Airport (KPIT) and Allegheny County Airport (KAGC) were used in the present work. Wind is highly dependent on local conditions and sensor mounting elevation so the airport values can provide only a rough estimate of local conditions, but both airports reported negligible wind of \SIrange{0}{5}{\meter\per\second} during the live fire test, so the effects of wind were not investigated.

\section{Description of Live Fire Tests}
ShotSpotter sensor arrays are routinely tested prior to active use by law enforcement. The live fire test of the ShotSpotter system examined here was conducted in Pittsburgh, PA by the Pittsburgh Bureau of Police in December, 2018. Pittsburgh has mixture of hilly and flat terrain. The regions covered by these tests primarily contain detached structures one to three stories in height, comparable to the scale of structures in many other American cities. Most sensors are installed on rooftops; the major component of the variance in sensor elevation is due to the hilly terrain. Several of the firing positions were adjacent to institutional buildings that generated loud, sharp echoes.

\begin{figure}[htpb]
\centering
\includegraphics[width={.45\textwidth}]{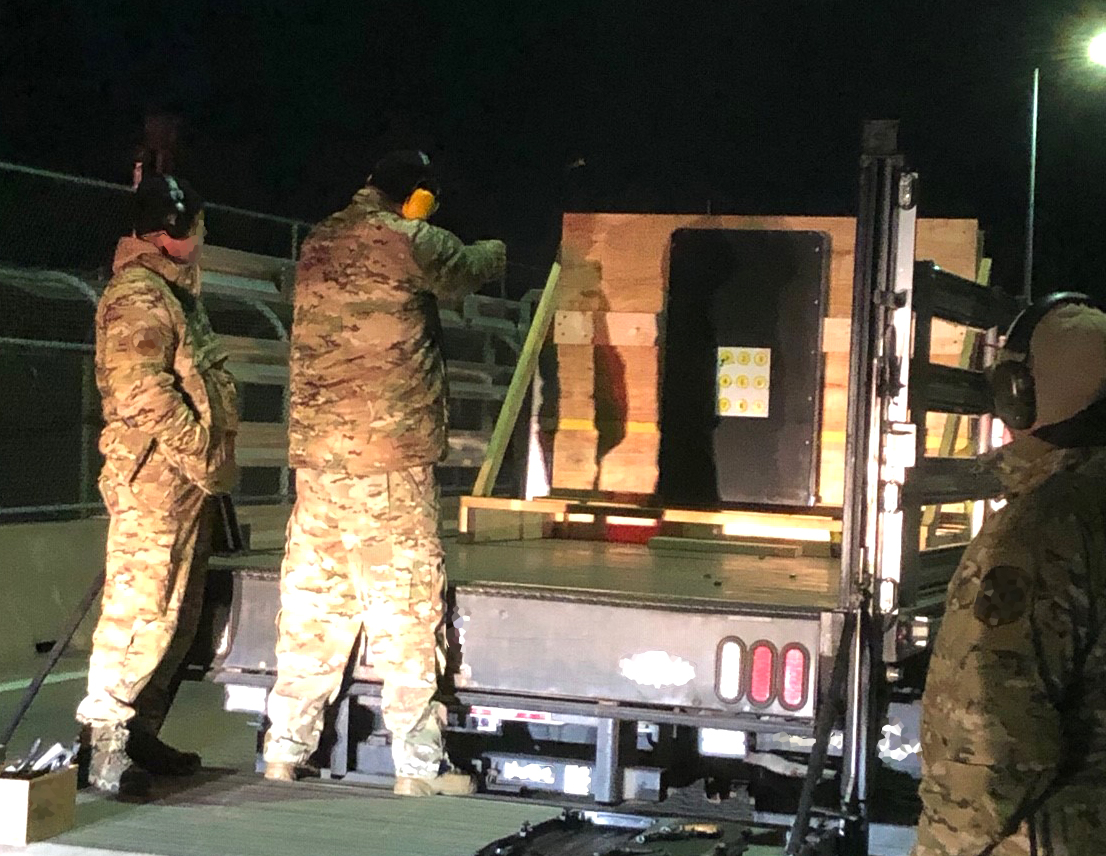}
\caption{Firing Position 1, showing the shooter and bullet trap, illustrating the source of the signals used for the test.}
\label{fig:FP1bullettrap}
\end{figure}

Officers secured the target area prior to each test for safety reasons. Live rounds were discharged into a bullet trap at distance $\approx$ \SI{3}{\meter} from the firearm, Figure (\ref{fig:FP1bullettrap}). All shots at a giving firing position were discharged with the muzzle of the firearm pointed in the same direction, which was not otherwise specified. The survey location of each firing position was determined by ShotSpotter personnel during the test using a smartphone application and qualitatively verified using aerial photographs available on Google Earth. Survey error $\epsilon$ was not well-characterized during the test but typical performance of such devices (which make use of signals from cellular and WiFi networks in addition to GPS) when used outdoors with good sky view is on the order of $\epsilon=$ \SI{2.5}{\meter}.

\begin{table}[htpb]
\begin{ruledtabular}

\centering
\begin{tabular}{llcccc}
FP & description & $N/M$ & $n$ & $\epsilon$ & $\rho$ \\
      &                   &             &       & (\si{\meter}) & (\si{{sensors}\per\kilo\meter\squared}) \\
\hline
FP1 & overpass \textasciitilde \SI{35}{\meter} above grade 
                                            & 36/36 & 17.8 & 4.14 & 9.4  \\
FP2 & parking lot near top of hill          & 36/36 & 14.3 & 5.20 & 11.4 \\
FP3 & intersection on high bluff            & 36/36 &  9.9 & 3.49 & 7.5  \\
FP4 & intersection near top of hill         & 36/36 & 13.8 & 4.76 & 9.4  \\
FP5 & playing field at head of valley       & 33/36 & 11.6 & 2.21 & 8.5 \\
FP6 & playing field near top of hill        & 36/36 & 13.6 & 5.54 & 8.5  \\
FP7 & playing field at end of valley        & 36/36 & 14.9 & 7.88 & 12.4 \\
FP8 & intersection at bottom of valley      & 36/36 & 11.2 & 2.18 & 9.9 \\
FP9 & parking lot on sloped region          & 36/36 & 10.6 & 6.93 & 8.0 \\

    \end{tabular}
    \caption{Realtime performance at the nine firing positions (FP) analyzed in present study. Buildings at most sites comprised detached residential structures 1--3 stories in height, with some larger commercial, residential, and institutional buildings. Ratio $N/M$ is number of shots $N$ detected out of shots fired $M$; $n$ is mean participating sensor count per shot; survey error $\epsilon$ is the distance from the acoustically-determined centroid to the survey location determined by smartphone-assisted GPS. Sensor density $\rho$ is in \si{{sensors}\per\kilo\meter\squared}. This is 2D performance using the algorithm matching the most sensors for each shot.}
    \label{tab:firingpoints}
\end{ruledtabular}
\end{table}

\begin{figure}
    \centering
\includegraphics[width={.45\textwidth}]{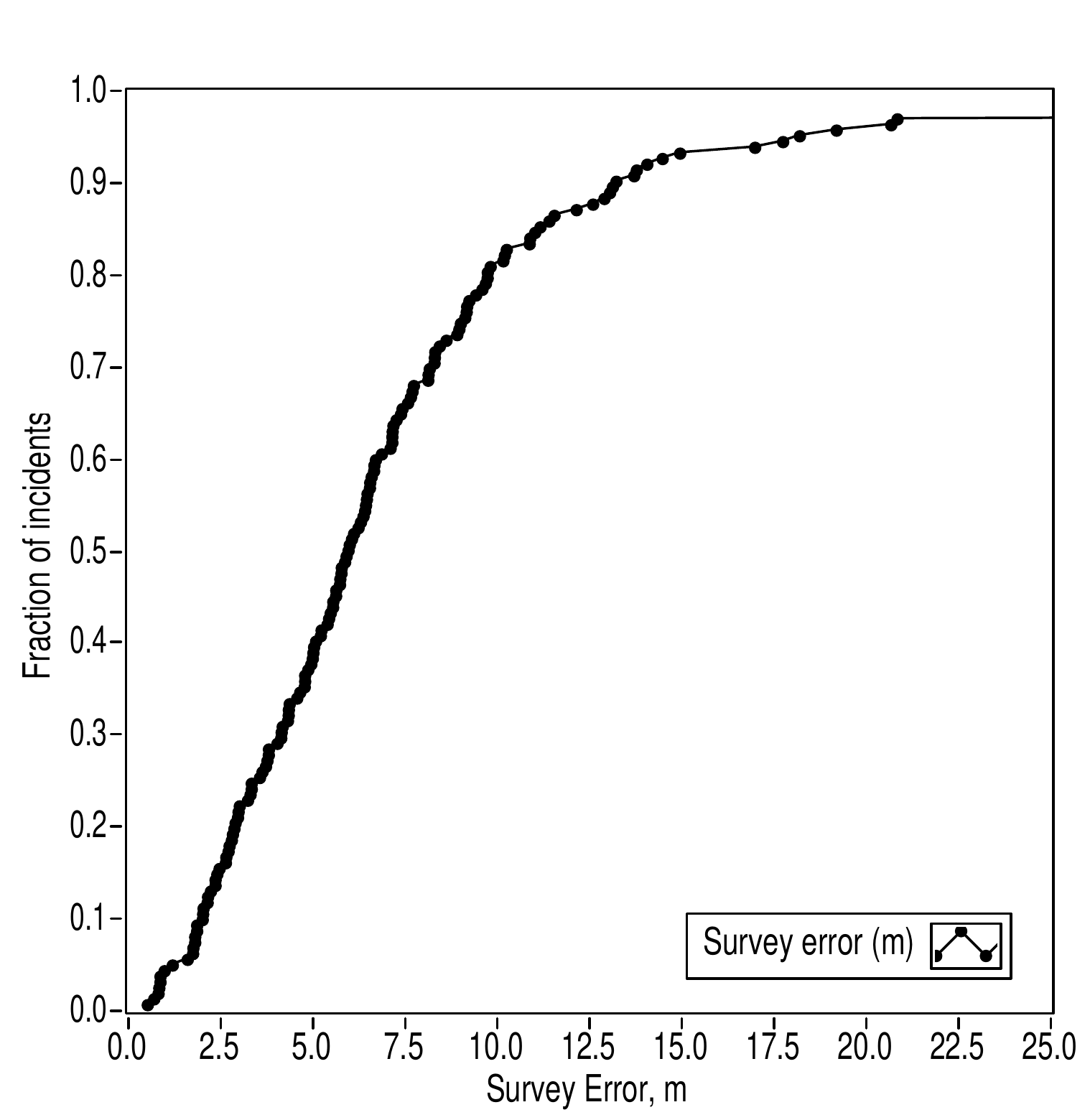}
    \caption{Cumulative distribution function of survey error during live-fire testings. For live fire incidents, 3-shot bursts are detected and scored as a single incident.}
    \label{fig:dqv-cdf}
\end{figure}

\begin{figure*}[htpb]
\centering
\includegraphics[width={.97\textwidth}]{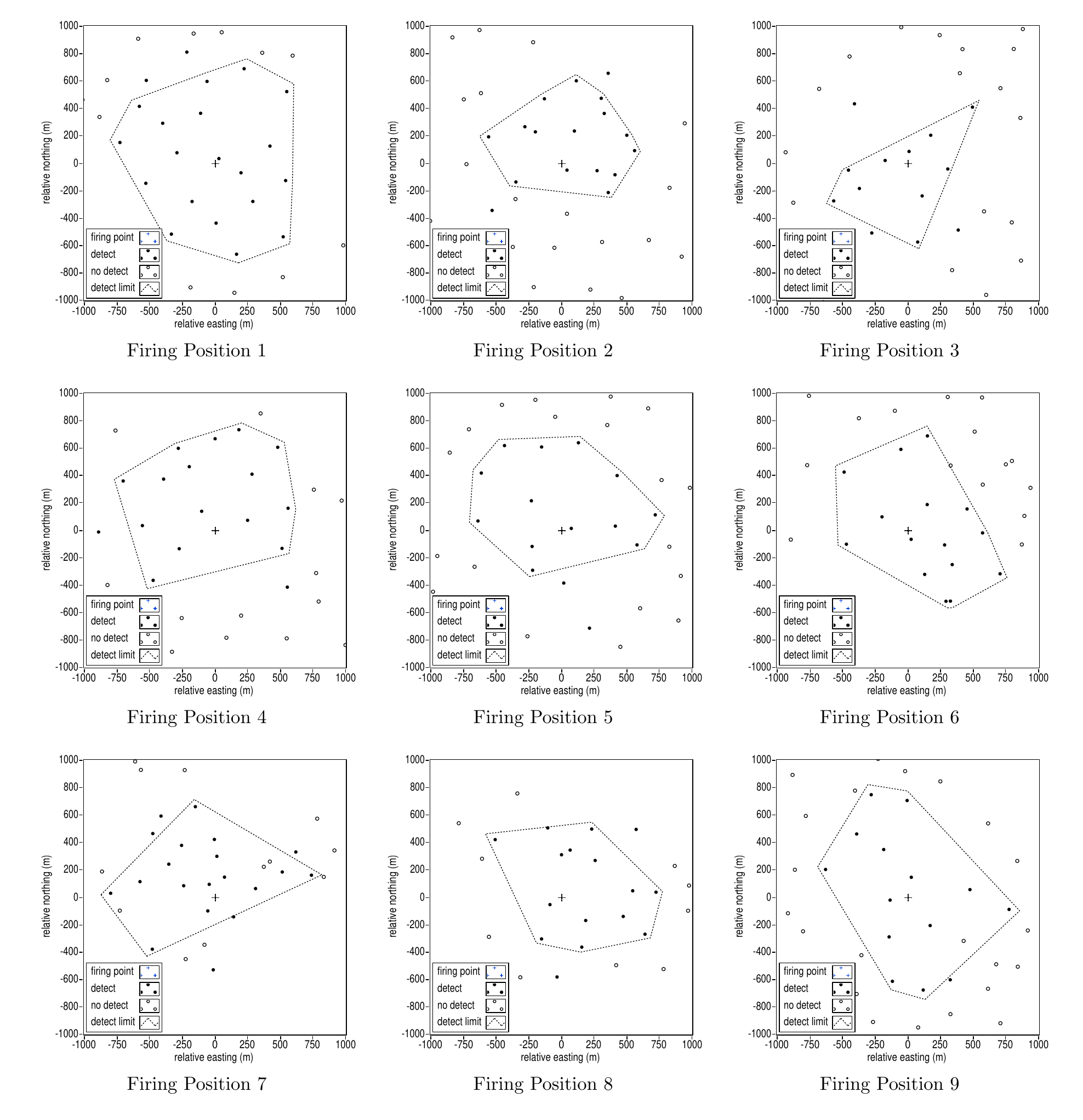}
\caption{Sensor positions shown relative to Firing Positions 1--9, which are located at the center of each plot. The filled circles show those sensors that participated in at least one shot; the dashed ``detection limit'' is the convex hull of those sensors reporting in at least 50\% of shots.}
\label{fig:convex-hulls}
\end{figure*}
Testing followed the standard firing sequence used by ShotSpotter. This comprises twelve shots (three single shots and three groups of three shots) for each firing position and tested weapon. This test used three different handguns at each firing position (.45 cal, .40 cal, \SI{9}{\milli\meter}). The test sequence was repeated at each of nine firing positions, for a total of 162 tests comprising 324 shots. A general description of each firing position is given in Table \ref{tab:firingpoints}. Temperature during the test was $-1 \pm 1 ^{\circ}\textrm{C}$ with light (0--2 m/s) wind. During the real-time tests, the system detected 100\% of incidents and located 96.9\% of incidents within a \SI{25}{\meter} benchmark criterion. The cumulative distribution function of survey error below \SI{25}{\meter} for all tests is shown Figure \ref{fig:dqv-cdf}. The real-time ShotSpotter system bundles multiple closely-spaced shots into a single incident, so scoring is per-test rather than per-shot.

Following each activation, the system requests two-second samples of 12 kHz audio to facilitate machine classification.  The results of the on-sensor pulse analysis are stored in a database for later use. Following the test, ShotSpotter personnel retrieved up to eight seconds of additional audio from selected sensors located within \SI{800}{m} of the firing position. The range limit is in alignment with the audio retrieval policy adopted by the company when servicing forensic requests. The present study limits performance analysis to sensors located within an \SI{800}{m} radius of each firing position.

Selected data from these tests are available as supplementary materials. These materials include pulse data (impulse arrival time, signal-to-noise ratio, acoustic power and sensor position) and up to six 12 kHz audio files for each test. Courts of law routinely redact sensor locations from trial records because of the risk of retaliation against property owners; accordingly, the sensor locations in the dataset are provided relative to an arbitrary (and unpublished) origin that is different for each firing position. 

The pulse data include a record of impulsive sounds triggered during the time range around each test. These include muzzle-blast sounds with near line-of-sight propagation paths that are useful for multilateration location as well as multipath signals and other impulsive noise sources that are unrelated to the gunfire test. ShotSpotter uses various pattern-matching and combinatorial consistency algorithms to pick sets of pulses suitable for multilateration. While these routines are not revealed in the present work, the output of the pulse selection algorithms (as tuples of pulses for each shot of each test) are provided as supplementary materials.

\section{Analysis}
\subsection{Multilateration Implementations}
The nine firing positions (Table \ref{tab:firingpoints}) were re-analyzed offline to evaluate the performance of the four 2D implementations available (Reddi, LeastSquares, MLG, and IDT). The input data were tuples of shortest acoustic path pulses within \SI{800}{\meter}. The mean number of participating sensors per shot ranged from 9.6 to 17.5, well over the four needed for unambiguous 2D multilateration. Results by implementation are shown in Table \ref{tab:algos}. The survey error $\epsilon$ is the two-dimensional distance from the acoustic multilateration centroid to the survey location determined by smartphone GPS; $\sigma_1$ and $\sigma_2$ are the square roots of the eigenvalues of the two-dimensional covariance matrix. Two-dimensional distance is considered the appropriate error measure in a public safety context. 

All four routines solve the same mathematical problem---minimizing the difference between the predicted and measured arrival times at the several sensors assuming a straight-line propagation model---so unsurprisingly similar results are obtained from all four algorithms. The Reddi implementation uses the first reporting sensor as a reference, which gives more weight to the nearest sensor than to all of the rest. The LeastSquares and related IDT routines yield good accuracy and stability, but the LeastSquares requires an extra TDoA and the IDT routine has a running time roughly ten times longer than Reddi or MLG for the same input data because it is iterative. The MLG implementation requires only the minimum number of TDoAs for a solution, does not arbitrarily overweight a single sensor, works in two or three dimensions and also has the fastest running time, since the computation required to set up $\mathbf{R}$ is simple and $\mathbf{R}^\dagger$ can be computed using SVD. The MLG implementation was used in the remainder of the study.

\begin{table}[htbp]
\begin{ruledtabular}
    \centering
    \begin{tabular}{clccccc}
 
Algo    & FP & $N/M$ & n & $\epsilon$ & $\sigma_1$ & $\sigma_2$ \\
         &   &    &   & $(\textrm{m})$ & $(\textrm{m})$  & $(\textrm{m})$ \\
\hline
Iterative & FP1 & 36/36 & 17.5 & 3.14 & 0.62 & 0.41 \\
Discharge & FP2 & 36/36 & 17.5 & 5.14 & 1.40 & 0.25 \\
Time & FP3 & 36/36 & 17.5 & 0.61 & 1.40 & 0.83 \\
(IDT) & FP4 & 36/36 & 17.5 & 4.09 & 2.33 & 0.56 \\
 & FP5 & 36/36 & 14.0 & 3.25 & 1.71 & 0.93 \\
 & FP6 & 36/36 & 14.0 & 5.21 & 1.29 & 0.40 \\
 & FP7 & 36/36 & 13.9 & 7.17 & 0.94 & 0.39 \\
 & FP8 & 36/36 & 14.0 & 2.67 & 1.39 & 0.47 \\
 & FP9 & 36/36 & 9.9 & 6.58 & 0.66 & 0.55 \\
\hline
Reddi & FP1 & 36/36 & 9.9 & 3.55 & 1.03 & 0.62 \\
 & FP2 & 36/36 & 9.8 & 4.66 & 1.05 & 0.33 \\
 & FP3 & 36/36 & 9.9 & 3.56 & 2.40 & 0.74 \\
 & FP4 & 36/36 & 13.8 & 7.69 & 5.52 & 1.86 \\
 & FP5 & 36/36 & 13.6 & 2.13 & 6.60 & 0.77 \\
 & FP6 & 36/36 & 11.2 & 6.30 & 0.76 & 0.50 \\
 & FP7 & 26/36 & 13.6 & 18.37 & 5.49 & 3.21 \\
 & FP8 & 36/36 & 11.4 & 2.15 & 1.39 & 0.32 \\
 & FP9 & 36/36 & 11.4 & 6.78 & 5.06 & 0.84 \\
\hline
LeastSquares & FP1 & 36/36 & 11.4 & 3.09 & 0.62 & 0.35 \\
 & FP2 & 36/36 & 11.4 & 4.01 & 1.68 & 0.20 \\
 & FP3 & 36/36 & 13.6 & 0.75 & 1.31 & 0.54 \\
 & FP4 & 31/36 & 13.6 & 3.93 & 1.35 & 0.71 \\
 & FP5 & 36/36 & 13.6 & 3.89 & 1.76 & 0.85 \\
 & FP6 & 36/36 & 13.6 & 5.62 & 1.47 & 0.50 \\
 & FP7 & 36/36 & 14.9 & 7.82 & 1.13 & 0.48 \\
 & FP8 & 35/36 & 9.6 & 2.48 & 1.95 & 1.60 \\
 & FP9 & 36/36 & 14.9 & 6.33 & 1.25 & 0.36 \\
\hline
MLG & FP1 & 36/36 & 14.9 & 2.94 & 0.55 & 0.42 \\
 & FP2 & 36/36 & 11.2 & 4.03 & 1.87 & 0.31 \\
 & FP3 & 36/36 & 11.2 & 0.29 & 1.49 & 0.23 \\
 & FP4 & 36/36 & 10.8 & 4.55 & 1.40 & 0.60 \\
 & FP5 & 36/36 & 11.2 & 4.20 & 1.97 & 0.67 \\
 & FP6 & 36/36 & 10.6 & 5.28 & 1.25 & 0.38 \\
 & FP7 & 36/36 & 10.5 & 7.86 & 0.96 & 0.37 \\
 & FP8 & 36/36 & 10.1 & 2.77 & 1.89 & 1.51 \\
 & FP9 & 36/36 & 10.6 & 5.76 & 0.91 & 0.56 \\

    \end{tabular}
    \caption{Performance of several 2D multilateration implementations at nine firing positions (FP). The ratio $N/M$ is number of shots $N$ detected out of $M$ shots fired, while $n$ is the mean number of sensors used in the location. The survey error $\epsilon$ is the 2D RMS distance from the survey location; $\sigma_1$ and $\sigma_2$ are the square roots of the eigenvalues of the covariance matrix, determined using PCA. These correspond to the standard deviations $\sigma_X$, $\sigma_Y$ of a bivariate normal distribution.}
    \label{tab:algos}
\end{ruledtabular}
\end{table}

\begin{figure*}[htpb]
\includegraphics[width={.97\textwidth}]{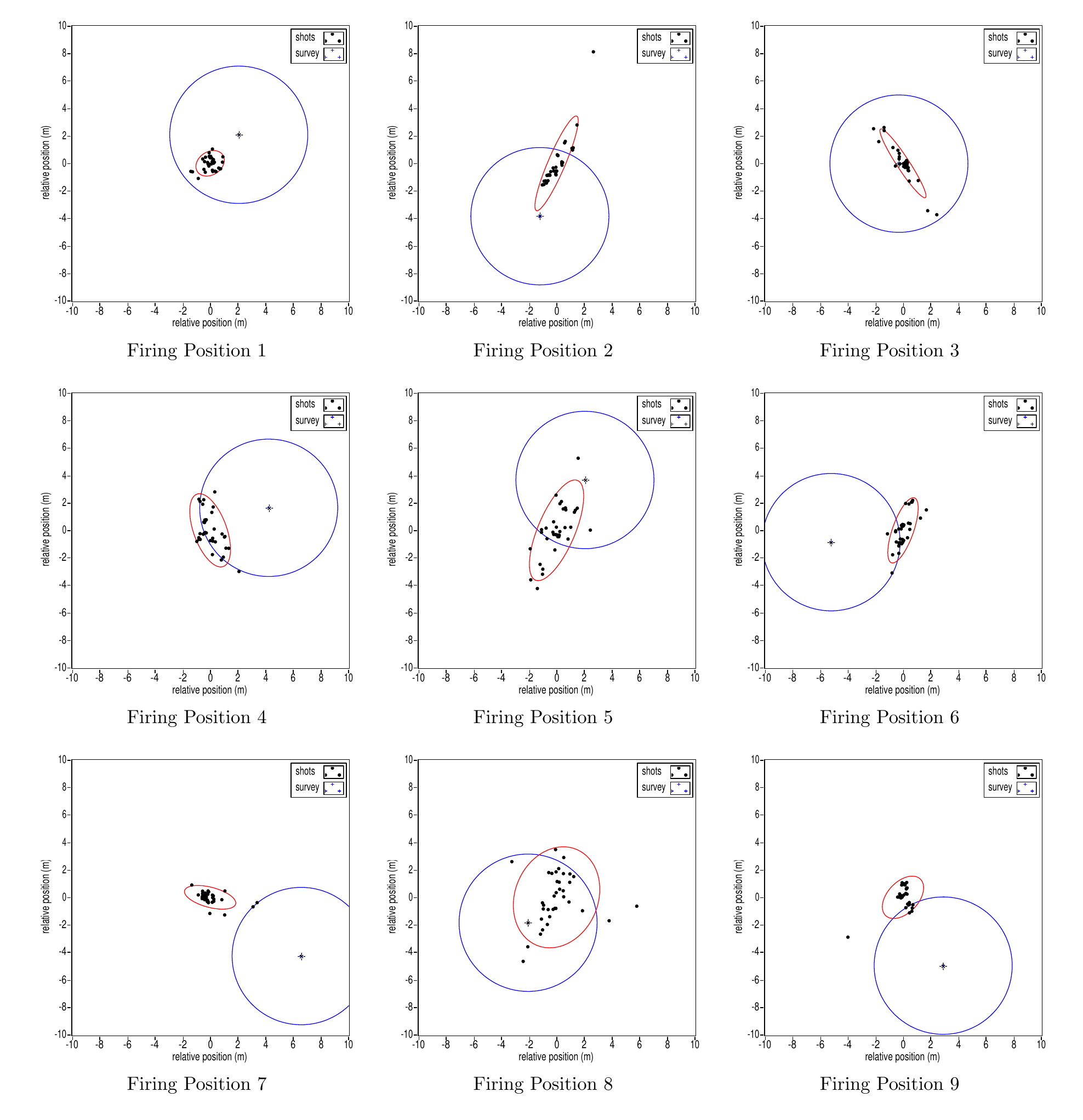}
\caption{Results of MLG implementation under 2D constraint for Firing Positions 1--9.}
\label{fig:MLG2D}
\end{figure*}

\subsection{Effects of geometric constraint}
While a ShotSpotter array in California's Central Valley has a vertical range of just 10 m over its entire 5 km extent, the Pittsburgh array under study has sensor elevation change (comprising terrain plus building height) of up to 160 m over 1 km. This makes the array suitable for a study of the effect of geometric constraint on location accuracy.

All incidents were relocated under the three geometric constraints (2D, 2.5D, 3D) to quantify the effect of including elevation data in the multilateration calculation. The system was permitted to use any algorithm in selecting pulse sets for multilateration; the MLG algorithm was used for the final location computation. The results (Table \ref{tab:constraints}) show that 2D location results are very similar, with no apparent improvement due to the use of 3D straight-line distance as a propagation model. 

A possible explanation is that a shot fired at ground level makes one diffracted path at the nearest building and then travels above the roofline along a path well-approximated by the 2D straight-line distance. 2D multilateration is implicitly locating in the plane of the sensors, so the system may be locating the best-fit location of a point slightly above the shooter. Because the 2D distances between sensors are so much larger than height of typical sensors above ground, the error introduced by this process is small.

These results suggest that the 2D geometric constraint is appropriate for acoustic multilateration of gunshots even in hilly terrain. The resulting locations are of high accuracy with precision of a few meters even when the sensors comprising the array contain a non-plane component. When a sufficient number of sensors are available for 3D multilateration, the results obtained show acceptable 2D location accuracy. However, the error in the $z-$axis is high, and the results show that the largely planar arrays used in this study do not provide sufficient elevation determination to identify the floor of a multi-story structure from which a shot might be fired.
\begin{figure}
    \centering
\includegraphics[width={.45\textwidth}]{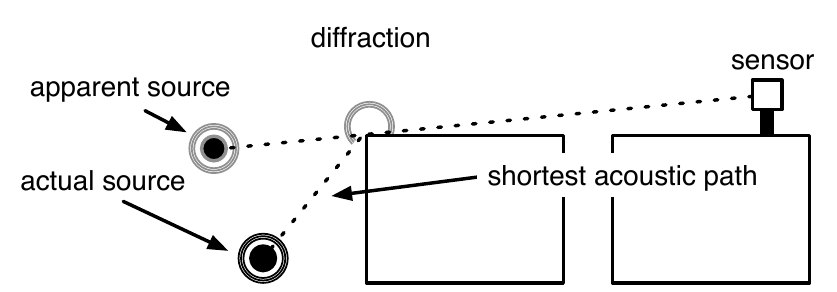}
    \caption{Schematic side view of how sound from a muzzle blast discharged near ground level relies on sound diffraction to reach a roof-mounted sensor. From the point of view of the sensor, the source appears to be an above-ground-level source with attenuated amplitude and nearly 2D propagation path.}
    \label{fig:diffraction1}
\end{figure}

\begin{figure}
    \centering
\includegraphics[width={.45\textwidth}]{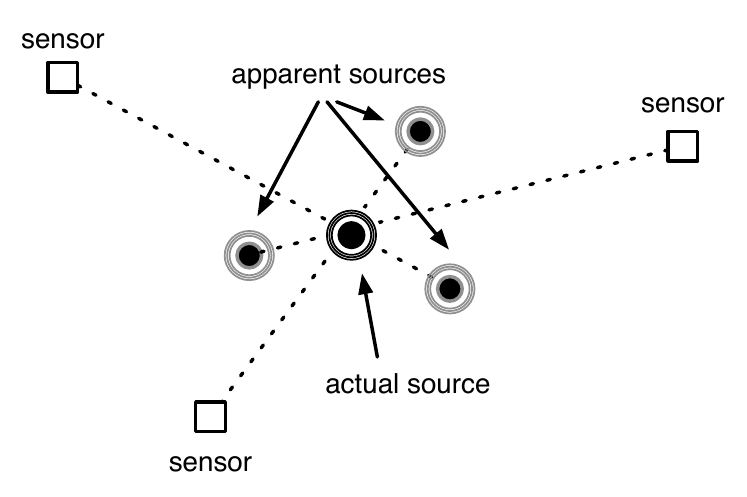} \\
    \caption{Schematic top view of the shortest acoustic path from a shot discharged near ground level to several rooftop-mounted sensors. Diffraction shifts the apparent position of the source from the point of view of each reporting sensor, but errors from multiple sensors tend to cancel when sensors surround the actual source.}
    \label{fig:diffraction2}
\end{figure}

Table \ref{tab:constraints} also shows the elevation survey error $\epsilon_z$, which is the RMS difference between presumed shooter elevation of \SI{1.5}{\meter} above grade and the output of the multilateration algorithm. The elevation survey error is only meaningful under the 3D constraint, as the shooter elevation is arbitrarily set to \SI{1.5}{\meter} above grade under the 2D and 2.5D geometric constraints. 

\begin{table}[htbp]
\begin{ruledtabular}
    \centering
    \begin{tabular}{llcccccc}
 & FP & $N/M$ & n & $\epsilon$ & $\sigma_1$ & $\sigma_2$  & $\epsilon_z$\\
 &  & & $(\textrm{m})$ & $(\textrm{m})$  & $(\textrm{m})$ & $(\textrm{m})$  & \\
 \hline
2D & FP1 & 36/36 & 17.53 & 2.96 & 0.67 & 0.28 & 34.95 \\
 & FP2 & 36/36 & 14.00 & 3.60 & 2.82 & 0.38 & 0.22 \\
 & FP3 & 36/36 & 9.86 & 0.36 & 1.43 & 0.58 & 0.54 \\
 & FP4 & 36/36 & 13.72 & 5.57 & 1.26 & 0.75 & 0.35 \\
 & FP5 & 36/36 & 11.44 & 3.30 & 2.18 & 0.76 & 0.00 \\
 & FP6 & 36/36 & 13.61 & 5.87 & 1.56 & 0.76 & 0.52 \\
 & FP7 & 36/36 & 14.94 & 4.57 & 0.48 & 0.23 & 0.37 \\
 & FP8 & 36/36 & 11.17 & 3.11 & 2.41 & 1.59 & 0.66 \\
 & FP9 & 36/36 & 10.58 & 6.44 & 1.14 & 0.46 & 0.71 \\
\hline
2.5D & FP1 & 36/36 & 17.53 & 3.02 & 0.70 & 0.23 & 34.95 \\
 & FP2 & 35/36 & 13.58 & 3.62 & 2.72 & 0.35 & 0.22 \\
 & FP3 & 36/36 & 9.83 & 0.54 & 1.21 & 0.38 & 0.54 \\
 & FP4 & 36/36 & 13.78 & 5.66 & 1.43 & 1.08 & 0.35 \\
 & FP5 & 36/36 & 11.44 & 1.96 & 1.87 & 0.79 & 0.00 \\
 & FP6 & 36/36 & 13.61 & 6.77 & 1.07 & 0.77 & 0.52 \\
 & FP7 & 36/36 & 14.94 & 4.91 & 0.85 & 0.44 & 0.37 \\
 & FP8 & 36/36 & 11.17 & 4.96 & 2.42 & 1.64 & 0.66 \\
 & FP9 & 36/36 & 10.58 & 6.15 & 1.15 & 0.32 & 0.71 \\
\hline
3D & FP1 & 36/36 & 17.53 & 3.76 & 0.51 & 0.32 & 42.40 \\
 & FP2 & 35/36 & 13.58 & 4.46 & 2.56 & 0.28 & 14.57 \\
 & FP3 & 36/36 & 9.83 & 2.51 & 1.95 & 0.72 & 45.55 \\
 & FP4 & 36/36 & 13.78 & 5.63 & 1.35 & 1.01 & 4.79 \\
 & FP5 & 36/36 & 11.44 & 2.29 & 3.17 & 0.73 & 3.82 \\
 & FP6 & 36/36 & 13.61 & 6.43 & 0.81 & 0.62 & 20.35 \\
 & FP7 & 36/36 & 14.94 & 4.84 & 1.73 & 0.81 & 3.73 \\
 & FP8 & 36/36 & 11.17 & 4.11 & 2.92 & 1.66 & 41.69 \\
 & FP9 & 35/36 & 10.28 & 5.68 & 0.86 & 0.38 & 18.75 \\
\hline

    \end{tabular}
    \caption{Performance of MLG multilateration implementation under 2D, 3D, and ``2.5D'' (3D constrain-to-plane) geometric constraints. Table columns have the same meaning as Table \ref{tab:algos}, with the addition of $\epsilon_z$, the distance from multilateration elevation to survey elevation. Note that for 2D and 2.5D solutions, the elevation coordinate is set arbitrarily at \SI{1.5}{\meter} above the ground elevation at the latitude, longitude of the multilateration results.}
\label{tab:constraints}
\end{ruledtabular}
\end{table}

\subsection{Effects of sensor density}
A \textit{reporting sensor} is one on which at least one gunshot impulse is detected programmatically by a sensor from shots fired at a particular location. A \textit{participating sensor} is one identified by the pulse selection algorithms as a member of a subset suitable for multilateration.

This series of firing tests was conducted on a night with still air after autumn leaf drop, resulting in good acoustic propagation of the gunshot sounds. The reporting sensor count was as high as 21 for the most powerful firearm (.45 cal) fired from the elevated position at FP1; the median number of participating sensors across all weapons and firing positions was 10, well above the minimum $d + 2 = 4$ participating sensors required for unambiguous 2D or 2.5D multilateration. 

To investigate the effects of reduced deployment density, the multilateration performance of reduced-density arrays was evaluated using a Monte-Carlo simulation. These simulations make use of the real-time pulse detection performance; the simulation is in varying the set of sensors considered to have been deployed during the test. 25 randomly-generated arrays were constructed at each firing point for the desired number of participating sensors. Simulations were run for 4, 5, 6, 8, 10, 12 and 15 sensors, corresponding to a sensor density of approximately \SIrange{2}{7.5}{{sensors}\per\kilo\meter\squared}. In some cases the target array density exceeded the number of sensors participating in a shot, so the maximum value of participating sensors was used. Multilateration was performed using the MLG algorithm under the 2D geometric constraint. Simulated incidents were filtered using the same pulse selection and filtering criteria used in production systems, with the result that some valid multilateration solutions were discarded because the signal amplitude on the strongest participating sensor failed to meet the (arbitrary) 73 dB SPL amplitude threshold applied to all incidents.

The results show that detection rates and location accuracy increase as the number of participating sensors increases. With six participating sensors, 96.3\% of incidents can be located to an accuracy of \SI{15}{\meter} or better. Detection rates and location accuracy continue to improve as the number of participating sensors increases above six, but additional reporting sensors provide only marginal improvements. The six-sensor result may be interpreted as ``sufficient to provide good array geometry after removing non-line-of-sight paths''.

As this technique varies the deployed sensor density rather than the propagation distance, it is best used to estimate the precision and accuracy obtainable as a function of the number of reporting sensors, a known quantity in a forensic context. It is less suitable for predicting the sensor density required to ensure good detection rates under non-ideal conditions. The presence of wind, rain, traffic noise and similar background noise sources reduce signal-to-noise ratio. Furthermore, while the simulation shows that any set of six reporting sensors will yield accurate multilateration results, implicit in the simulation is that the reporting sensors are \textit{well-distributed}, since the sensors were randomly drawn from a set of participating sensors that fully surround the shooter, as shown in Figure \ref{fig:convex-hulls}. Wind may refract the sound away from a given direction entirely, and some barriers (such as highway embankments or sound walls) may attenuate the sound below the point of detectability, potentially resulting in a situation where the shooter is far outside the convex hull of reporting sensors. Because the effect is due to the geometry of the receiver array, the error of a multilateration solution can be estimated using the analytical approach of \citep{mathias2008efficient} or using Monte-Carlo simulation, in which the shooter location is repeatedly re-computed after perturbing the input measurements according to the measurement error estimates in Section \ref{sysdesign}. Because of the high number of participating sensors in the Pittsburgh live-fire test, the data published in the present work provide a good basis for future study of the effects of array geometry on location accuracy.

\begin{figure*}[htpb]
\centering
\includegraphics[width={.97\textwidth}]{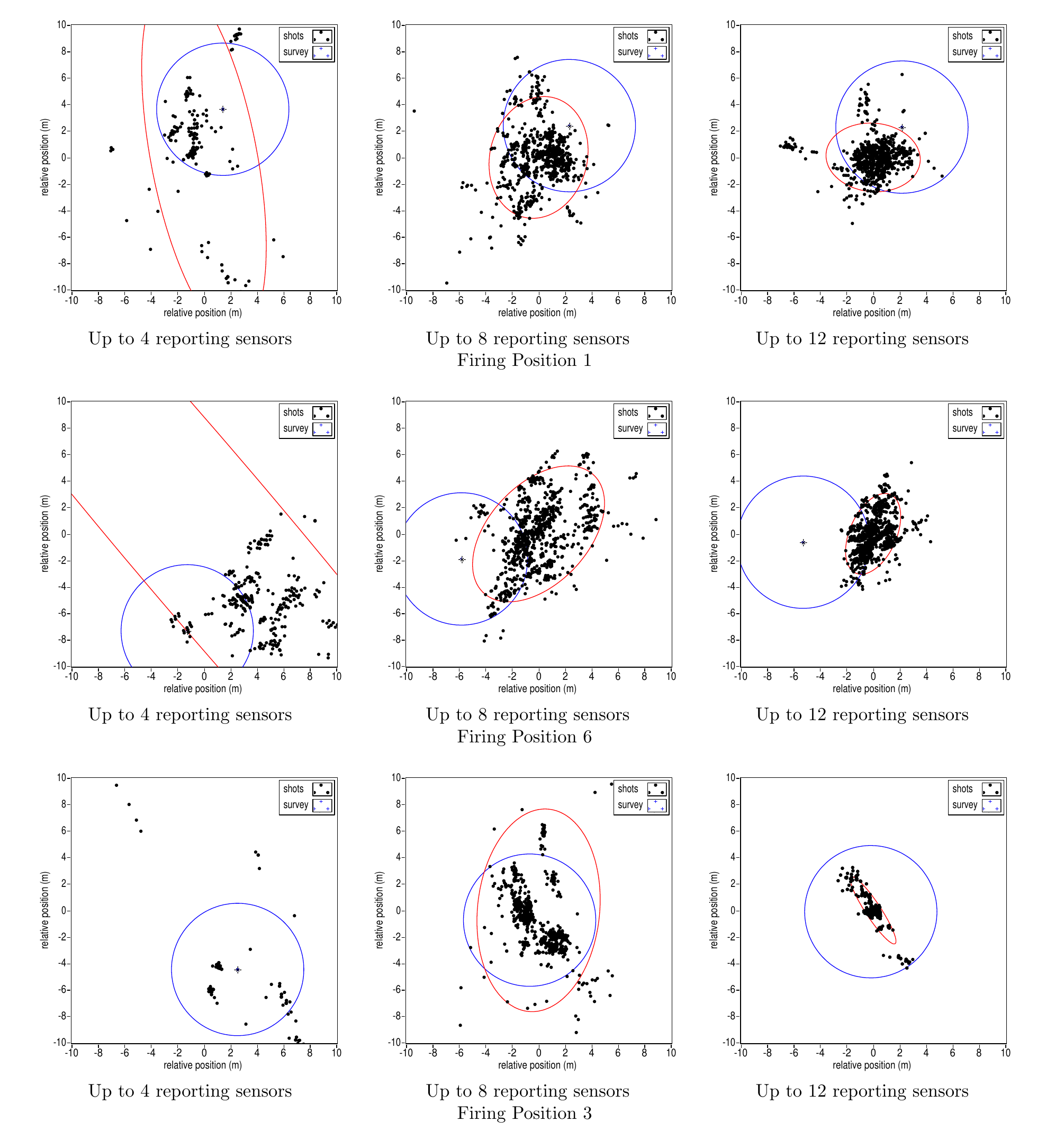}
\caption{Location accuracy of randomly-generated, reduced-density arrays, determined using a Monte-Carlo simulation of 25 random arrays of participating sensors at each firing position. These plots show performance with a maximum of 4 (left column), 8 (middle column), or 12 (right columns) sensors, corresponding to a sensor density of \SIrange{2}{6}{{sensors}\per\kilo\meter\squared}. Firing positions shown reflect those with highest sensor participation during the live fire test (FP1, top row), medium participation (FP6, middle row) or low participation (FP3, bottom). See \ref{fig:sensordensity2} for detection rates and location accuracy at all firing positions. These results show that the robustness of multilateration results is improved by deploying a sufficient number of sensors to ensure good participation.}
\label{fig:sensordensity}
\end{figure*}

\begin{figure}[htpb]
\centering
\includegraphics[width={.45\textwidth}]{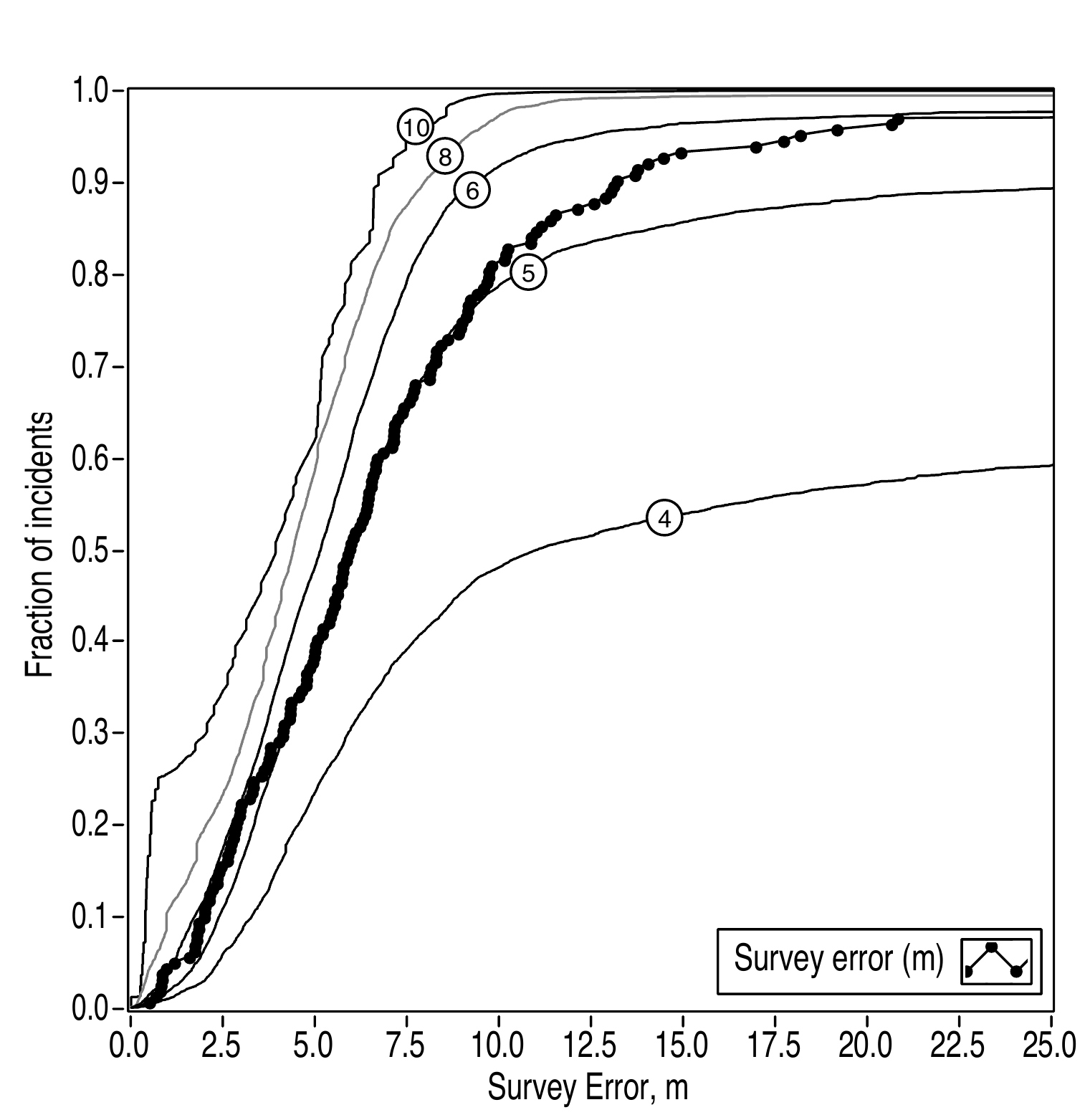}
\caption{Cumulative distribution function (CDF) of simulated detection accuracy with random arrays containing varying numbers of participating sensors. The CDF for each simulations is based on relocating all shots using 25 random arrays comprising 4, 5, 6, 8, or 10 participating sensors at each firing position, corresponding to a sensor density of approximately \SIrange{2}{5}{{sensors}\per\kilo\meter\squared}. The CDF of the live-fire test from Figure \ref{fig:dqv-cdf} is included for comparison.}
\label{fig:sensordensity2}
\end{figure}

\section{Conclusion}
Source location via acoustic multilateration is a well-developed technique that was reduced to practice during the First World War. Inexpensive computation and cellular radio networks allow the use of these techniques by automated systems to detect gunshots in near real time. 

For the sensor arrays considered, the multilateration algorithm of \citep{mathias2008efficient} had the greatest precision. Use of a 3D straight-line propagation model (either unconstrained or constrained to the ground plane) did not significantly improve location accuracy over the simpler 2D propagation model.  Under these conditions, acoustic multilateration provided a 2D accuracy between \SIrange{0.36}{6.44}{\meter} with respect to surveyed location and a 2D precision (or relative accuracy) of \SIrange{0.48}{2.82}{\meter}. High relative accuracy is essential when performing forensic analysis of situations that may involve multiple shooters. The precision of elevation estimates was much lower, with full 3D solutions yielding an RMS accuracy vs digital elevation model of \SI{17.8}{\meter}. The array geometry tested was not adequate to identify the elevation of shots fired from the upper stories of a building. In coverage areas where firearm discharge from an elevated structure is of concern, sensors should be positioned at both the roof and base of tall structures in the area to provide better elevation precision.

Detection on the minimum number of sensors required (nominally three in two dimensions) can produce an accurate result when acoustic paths are near line-of-sight, but the fully-determined solution is not robust under the presence of timing errors. Additional information (such as waveform analysis) must be used to estimate the likelihood that the TDoAs are good estimates of the line-of-sight travel time. With four sensors, the presence of a timing error can be detected (since each 3-tuple produces a different solution) but not corrected. With five or more sensors, error-inducing multipath inputs may be identified and rejected. Simulations with random sensor arrays show that excellent precision and accuracy are reliably obtained with arbitrary arrays of six reporting sensors drawn at random from a spatially-distributed set of participating sensors. Detection rate and accuracy continue to improve as the number of participating sensors increases, with additional sensors above six providing only marginal improvement. The six-sensor criterion corresponds to a deployed sensor density of approximately \SI{3}{{sensors}\per\kilo\meter\squared} in the arrays studied; a higher sensor density is needed to ensure sufficient sensor participation in practical deployments. Higher density arrays help ensure an adequate participating sensor count (preferably six, but at least three) in highly built-up areas, at the edge of coverage areas, or in adverse weather conditions. The ideal sensor deployment configuration is one that will detect a high fraction of outdoor firearm discharges from any part of the coverage area under all weather conditions, and do so using a minimal number of sensors. A model that incorporates the combined effects of structures, terrain, foliage, wind, and ground reflection would be of value in the design of effective, economical sensor arrays for acoustic gunshot location, and is a topic we seek to address in future work.

\begin{acknowledgments}
The authors thank the Pittsburgh Bureau of Police for conducting a comprehensive live-fire test of the ShotSpotter system in their community. We also thank the ShotSpotter Operations Team, and Project Manager Athnaiel Bazi in particular, for designing, deploying, and maintaining an effective array for gunshot detection.

The authors are employed by and hold an ownership interest in ShotSpotter, Inc.
\end{acknowledgments}



\end{document}